\documentclass[acmsmall,screen,nonacm]{acmart}

\usepackage{booktabs}
\usepackage{graphicx}
\usepackage{multirow}
\usepackage{subcaption}
\usepackage{algorithm}
\usepackage{algorithmic}
\usepackage{enumitem}
\usepackage{subcaption}
\usepackage{multicol}
\usepackage{multirow}
\usepackage{xcolor}
\usepackage{flafter}
\usepackage{placeins}
\usepackage{tabularx}
\usepackage{ragged2e}
\usepackage{pifont}

\definecolor{correct_green}{RGB}{40, 160, 40}
\definecolor{bar_blue}{RGB}{200, 220, 240}
\definecolor{text_gray}{RGB}{120, 120, 120}

\newcommand{\cmark}{\textcolor{correct_green}{\ding{51}}}

\newcommand{\method}{CaLIR}

\newcommand{\preditem}[3]{
    \makebox[0pt][l]{\color{bar_blue}\rule[-2pt]{#1\linewidth}{10pt}}
    \ifnum#3=1 
        \textbf{#2} \hfill \textbf{#1} \cmark
    \else 
        \textcolor{text_gray}{#2} \hfill \textcolor{text_gray}{\scriptsize #1} \phantom{\cmark}
    \fi
}

\setlist[itemize]{leftmargin=*}
\AtBeginDocument{
  }

\setcopyright{none}
\makeatletter
\AtBeginDocument{
  \fancypagestyle{firstpagestyle}{
    \fancyhf{}

    \fancyfoot[L]{\footnotesize Under Review}
    \fancyfoot[R]{\footnotesize\thepage}
  }
  \fancypagestyle{standardpagestyle}{
    \fancyhf{}

    \fancyhead[LE]{\@headfootfont\thepage}
    \fancyhead[RO]{\@headfootfont\thepage}
    \fancyhead[RE]{\@headfootfont\@shortauthors}
    \fancyhead[LO]{\@headfootfont\shorttitle}
    \fancyfoot[L]{\footnotesize Under Review}
  }
  \pagestyle{standardpagestyle}
}
\makeatother

\begin{document}

\title{Beyond Matching: Category-Guided Latent Intent Reasoning for Generative Retrieval in E-Commerce}

\author{Fuwei Zhang}
\affiliation{
  \institution{Institute of Artificial Intelligence, Beihang University}
  \country{China}}
\affiliation{
  \institution{Meituan}
  \country{China}}

\author{Xiaoyu Liu}
\affiliation{
  \institution{Institute of Artificial Intelligence, Beihang University}
  \country{China}}

\author{Jiajie Jin}
\affiliation{
  \institution{Gaoling School of Artificial Intelligence, Renmin University of China}
  \country{China}}

\author{Jiale Mao}
\affiliation{
  \institution{Institute of Artificial Intelligence, Beihang University}
  \country{China}}
\affiliation{
  \institution{Meituan}
  \country{China}}

\author{Wei Chen}
\affiliation{
  \institution{Institute of Artificial Intelligence, Beihang University}
  \country{China}}

\author{Dongbo Xi}
\affiliation{
  \institution{Meituan}
  \country{China}}

\author{Yifan Yang}
\affiliation{
  \institution{Meituan}
  \country{China}}

\author{Peng Yan}
\affiliation{
  \institution{Meituan}
  \country{China}}

\author{Zichao Hao}
\affiliation{
  \institution{Beijing Information Science and Technology University}
  \country{China}}

\author{Zhao Zhang}
\authornote{Corresponding authors.}
\affiliation{
  \institution{SKLCCSE, School of Computer Science and Engineering, Beihang University}
  \country{China}}

\author{Fuzhen Zhuang}
\authornotemark[1]
\affiliation{
  \institution{Institute of Artificial Intelligence, Beihang University}
  \country{China}}

\renewcommand{\shortauthors}{Zhang et al.}

\begin{abstract}
Generative retrieval offers a new paradigm for e-commerce search by mapping user queries directly to product Semantic Identifiers (SIDs). However, e-commerce queries are often short, noisy, attribute-heavy, and associated with multiple category-consistent products, creating a substantial representation gap between natural-language shopping intent and artificially constructed item SIDs. Explicit Chain-of-Thought (CoT) reasoning can help bridge this gap, but its extra generation cost is difficult to reconcile with the low-latency requirements of online e-commerce systems. To address this challenge, we propose \method{} (Category-guided Latent Intent Reasoning), a category-guided latent intent reasoning framework for e-commerce generative retrieval. Rather than generating explicit textual rationales, \method{} learns continuous latent intent states before SID decoding and uses product category hierarchies as a natural scaffold for coarse-to-fine intent reasoning. Specifically, we introduce hierarchical semantic reasoning to align latent states with category-level shopping intent, and query-wise reasoning enhancement to model diverse intent paths under multi-positive queries. \method{} further combines a query-specific dynamic prefix trie, assembled from pre-indexed category-level tries, with reasoning-aware constrained decoding. Experiments on multilingual e-commerce search datasets show that \method{} achieves a better balance between retrieval effectiveness and inference efficiency than existing methods, while also demonstrating transferability and robustness across induced hierarchies and different generative backbones.
\end{abstract}

\ccsdesc[500]{Information systems~Information retrieval}
\keywords{E-Commerce Search, Generative Retrieval, Latent Reasoning, Category-Guided Reasoning, Information Retrieval}

\maketitle
\begingroup
\renewcommand{\thefootnote}{}
\footnotetext{Fuwei Zhang and Jiale Mao were interns at Meituan. This work was completed during their internships at Meituan. Faculty advisors from Beihang University provided remote guidance only.}
\endgroup

\section{Introduction}

\begin{figure}[t]
    \centering

    \includegraphics[width=\linewidth]{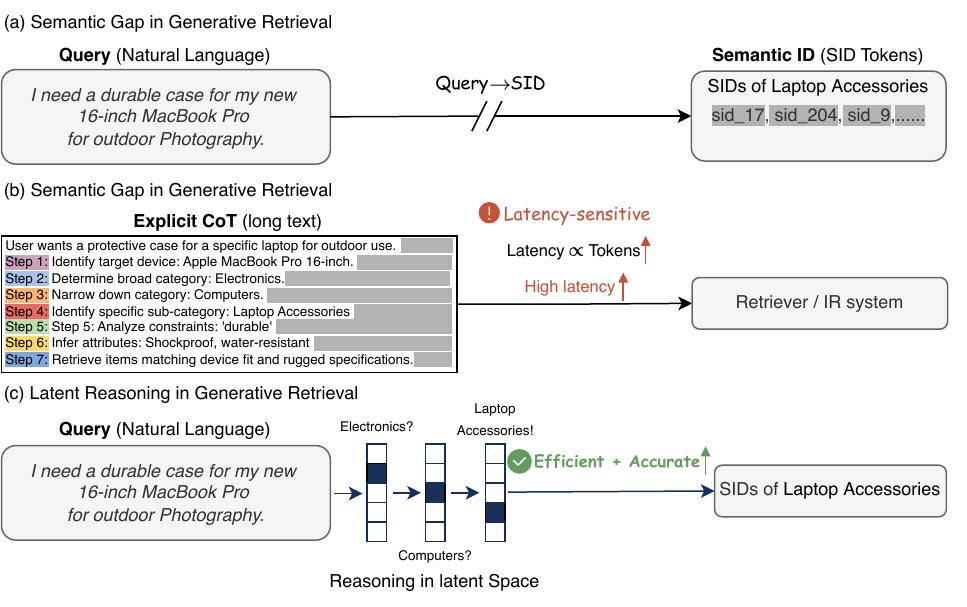}
    \caption{Comparison of Direct Generation, Explicit CoT, and Category-Guided Latent Intent Reasoning.}
    \label{fig:intro}
\end{figure}

E-commerce search is a central application of information retrieval, where a system must match user queries to relevant products in a large, dynamic catalog. Compared with open-domain web or document search, e-commerce search has several domain-specific characteristics: queries are usually short and intent-dense, product descriptions contain structured attributes such as brand, color, size, and function, and a single query may correspond to multiple valid products across related categories. These properties make query-product matching particularly sensitive to long-tail vocabulary, attribute constraints, negation, and category ambiguity.

Traditional e-commerce search systems primarily rely on sparse retrieval paradigms~\cite{bm25}, representing products and queries as high-dimensional word-frequency vectors. However, these methods often struggle to bridge the ``lexical gap'' between different expressions of the same shopping intent. While the subsequent dense retrieval~\cite{dense1,dense3} approach mitigates semantic mismatch, its robustness is still limited by the precision of vector embeddings. Specifically, dense retrieval struggles to effectively capture discriminative low-frequency product attributes and fine-grained constraints in long-tail catalog distributions.

Recently, the rapid advancement of generative language models has introduced a promising new paradigm for e-commerce retrieval: Generative Retrieval (GR). Distinct from traditional ranking-based approaches, GR leverages the deep semantic understanding and generation capabilities of these models to reformulate retrieval as a sequence-to-sequence task. By utilizing Semantic IDs (SIDs), which are typically constructed via clustering or quantization, this approach directly generates item identifiers from shopping queries.

However, a significant ``representation gap'' persists in current GR frameworks for e-commerce search.
As illustrated in Fig.~\ref{fig:intro} (a), most item SIDs are essentially abstract tokens (e.g., \texttt{sid\_17}) lacking inherent product semantics.
This creates a natural discrepancy between these tokens and natural language shopping queries.
Consequently, the model is forced to learn a direct mapping from queries to these ``foreign tokens'' during training, without explicitly modeling whether the query first points to a broad department, a product family, or a fine-grained leaf category.
While explicit Chain-of-Thought (CoT) prompting can bridge this gap through step-by-step reasoning (Fig.~\ref{fig:intro} (b)), the resulting high latency makes it impractical for online e-commerce search, where retrieval must respond under strict latency constraints.
To balance computational efficiency with intent grounding, Fig.~\ref{fig:intro} (c) introduces category-guided latent intent reasoning, which operates in a coarse-to-fine manner within the continuous latent space, first capturing coarse shopping-intent information and subsequently refining it into detailed product categories.

To address these challenges, we propose \textbf{\method{}} (\textbf{Ca}tegory-guided \textbf{L}atent \textbf{I}ntent \textbf{R}easoning), a framework for e-commerce search.
Adopting a coarse-to-fine architecture, \method{} incorporates shopping-intent reasoning within a continuous vector space.
This design effectively bridges the semantic gap between explicit shopping queries and item SID tokens while maintaining comparable inference efficiency. Specifically, \method{} diverges from standard explicit CoT by avoiding the generation of concrete text tokens.
Instead, it leverages the multi-level product categories naturally available in e-commerce scenarios as latent intent scaffolds, providing hierarchical semantic support before SID generation.
To facilitate the learning of category-guided latent intent reasoning, we design two supervised tasks:
1) \textbf{Hierarchical Semantic Reasoning}, which transforms the query into a coarse-to-fine category-level intent mapping. 2) \textbf{Query-wise Reasoning Enhancement}, which aligns category information associated with the same query to guide the latent intent reasoning process, thereby improving the diversity and robustness of reasoning paths for multi-positive e-commerce intents.
During the inference phase, we further introduce a \textbf{Reasoning-aware Constrained Decoding} strategy, which utilizes inferred intent categories to dynamically assemble a query-specific prefix tree from pre-indexed category-level tries for constrained beam search. This approach implements a ``reasoning-then-decoding'' pipeline to enhance e-commerce retrieval performance.
Here, we summarize our contributions:
\begin{itemize}
    \item We introduce category-guided latent intent reasoning into e-commerce GR by proposing \textbf{\method{}}. This framework bridges the ``representation gap'' between shopping queries and item SIDs within a continuous vector space, achieving a superior trade-off between intent grounding and inference efficiency compared to explicit Chain-of-Thought approaches.
    \item We develop a coarse-to-fine learning paradigm tailored to e-commerce catalogs, where latent intent paths are derived from hierarchical product categories. Specifically, we design \textbf{Hierarchical Semantic Reasoning} and \textbf{Query-wise Reasoning Enhancement} objectives to align latent states with category hierarchies and multi-positive shopping intents.
    \item We propose a \textbf{Reasoning-aware Constrained Decoding} strategy that leverages inferred intent categories to assemble query-specific dynamic trie constraints, narrow the catalog search space, and improve precise SID generation.
    \item We conduct extensive experiments on multilingual e-commerce search datasets, together with transferability and backbone analyses. The results demonstrate that \method{} significantly outperforms state-of-the-art GR methods, validating the effectiveness of category-guided latent intent reasoning in the main e-commerce search setting.
\end{itemize}

\section{Related Work}
\subsection{Sparse and Dense Retrieval}
Product search commonly builds on sparse and dense retrieval. Sparse methods, such as BM25, rely on statistical term-frequency for keyword matching and remain strong when exact product attributes are important. Conversely, dense retrieval paradigms like DPR \cite{dense1} and ANCE \cite{dense3} utilize neural embeddings to capture semantic similarity beyond surface-form overlap.
To unify these paradigms, hybrid models like COIL \cite{coil} have been developed. Furthermore, the Transformer \cite{vaswani2017attention} architecture has revolutionized the field; frameworks like Sentence-Transformers \cite{reimers2019sentence} leverage self-attention to produce high-quality sentence-level embeddings. Recent advancements continue to refine the intersection of statistical precision and neural semantic depth through learned sparse representations \cite{learnedsparse}. For multilingual retrieval, hierarchical knowledge enhancement has also been used to bridge cross-lingual query-document gaps~\cite{zhang2022mind}. Notably, BGE-M3 \cite{chen2024m3} stands out as a versatile text embedding model that unifies dense, sparse, and multi-vector retrieval functionalities within a single framework while supporting multi-lingual and long-text product encoding. These methods are important baselines for e-commerce search, but they still rely on retrieval-and-ranking pipelines rather than directly generating catalog item identifiers.

\subsection{Generative Retrieval}
Generative Retrieval (GR) reformulates retrieval as a sequence-to-sequence generation task, where a model directly predicts a unique identifier given a query~\cite{kuo2024survey,li2025matching}. In e-commerce search, this identifier corresponds to a product item or item group in a catalog, and the core challenge lies in constructing SIDs that ensure uniqueness while preserving semantic and hierarchical product structures.
Early milestones, such as DSI~\cite{dsi}, explore various identifier types, including atomic, naive, and semantic IDs. Subsequent architectures like DSI-QG~\cite{dsiqg}, SE-DSI~\cite{sedsi}, and DE-DSI~\cite{neague2024dsi} introduce further refinements to these mechanisms. Alternative approaches adopt text-centric identifiers~\cite{li-etal-2023-multiview,seal}. For instance, GENRE~\cite{genre} utilizes text-based identifiers, while SEAL~\cite{seal} leverages explicit substrings indexed by FM-indexes~\cite{fmindex}. Building on this, Ultron~\cite{zhou2022ultron} and D2-DocID~\cite{cheng2025descriptive} advance the field by introducing learnable, descriptive identifiers. NCI~\cite{nci} leverages neural architectures to improve retrieval effectiveness.
Learned tokenization has also become an important direction: GenRet~\cite{genret} learns document tokenization for GR, while MVDR~\cite{mvdr} shows that GR can be interpreted through the lens of multi-vector dense retrieval, connecting sequence generation with dense retrieval behavior.
Recent research pivots toward enhancing the hierarchical structure and semantic alignment of identifiers~\cite{li2024distillation,hi-gen,seater}. Methods such as Hi-gen~\cite{hi-gen} and SEATER~\cite{seater} employ K-means clustering and constrained K-ary trees to preserve structural relationships, with SEATER explicitly incorporating alignment loss. HierGR~\cite{zhang2025hiergr} further refines residual quantization to obtain hierarchical semantic IDs for food delivery search and discusses online deployment under scale, latency, and location constraints. To integrate e-commerce domain knowledge, CAT-ID$^2$~\cite{catid} embeds multi-level category information into the identifier learning process, ensuring identifiers reflect the underlying product category distribution. Similarly, MERGE~\cite{merge} introduces multi-level relevance alignment during the Semantic ID (SID) learning phase to enhance representation capability. In the adjacent generative recommendation setting, MACRec~\cite{zhang2026multi} incorporates multimodal information into both semantic ID learning and generative model training. A model-scaling analysis of SID-based generative recommendation~\cite{liu2025understanding} further suggests that SID capacity can become a bottleneck as model scale changes.
Beyond structure, recent works focus on ranking, scalability, corpus dynamics, constrained decoding, and agentic retrieval. LTRGR~\cite{ltrgr} and GR$^2$~\cite{grgr} incorporate ranking signals and multi-graded relevance through constrained contrastive training. ZeroGR~\cite{sun2025zerogr} studies zero-shot GR with automatically generated supervision, while dynamic-corpus studies~\cite{zhang2025replication} and model-editing methods~\cite{zhang2026model} examine how GR systems handle newly added documents without full retraining. A theoretical analysis of constrained auto-regressive decoding~\cite{wu2025constrained} further shows that corpus-specific constraints and beam search can limit GR generalization and top-$k$ recall. Agent-oriented IR work~\cite{zhang2024ai} highlights a generation-and-ranking view of retrieval systems. Finally, to address scalability and resource constraints, optimization frameworks such as RIPOR~\cite{ripor} and GDR~\cite{gdr} are proposed to improve the memory efficiency of GR. These studies mainly strengthen the identifier side of GR, including SID construction, SID alignment, ranking-aware training, multimodal SID learning, corpus updates, constrained decoding, and scalable retrieval. Our work is complementary: it focuses on the query-to-SID transition and learns category-guided latent intent states before identifier generation, so that decoding is grounded by coarse-to-fine shopping intent rather than by direct query-to-SID matching alone.

\subsection{Latent Reasoning}
Reasoning in Large Language Models (LLMs) has evolved from Explicit Chain-of-Thought (CoT) \cite{cot,liu2025onerec}, which is often bottlenecked by token-level decoding overhead, toward latent reasoning paradigms. Recent advances focus on internalizing rationales within the latent space: \textit{Coconut} \cite{coconut} and \textit{Quiet-STaR} \cite{zelikmanquiet} employ continuous latent vectors to enable ``thinking before speaking,'' while \textit{LLaDA} \cite{llada} utilizes masked diffusion for iterative refinement of latent trajectories. In recommendation, related efforts have begun to explore reasoning-oriented designs, including generative recommendation studies \cite{cao2025onepiece} and sequential recommendation methods such as \textit{Think before Recommend} \cite{tang2025think}, \textit{STREAM-Rec} \cite{zhang2025slow}, \textit{LARES} \cite{liu2025lares}, and \textit{LASAR}~\cite{chen2026lasar}. Specifically, \textit{STREAM-Rec} introduces a slow-thinking generative recommender that constructs intermediate reasoning tokens through residual-based iterative inference and then improves the reasoning process through supervised fine-tuning and reinforcement learning. \textit{LARES} instead performs depth-recurrent latent reasoning by repeatedly refining all input-token representations and trains this process with self-supervised trajectory- and step-level alignment followed by reinforcement post-training. \textit{LASAR} aligns latent reasoning trajectories with explicit CoT semantic anchors and learns adaptive reasoning depth for generative recommendation. These methods show that reasoning processes can improve sequential user-preference modeling, but they mainly focus on next-item recommendation rather than retrieval-oriented SID generation.

In the e-commerce Generative Retrieval (GR) scenario, however, latent reasoning remains under-explored despite its potential to bridge the gap between shopping-intent understanding and SID generation. Different from general latent reasoning methods that primarily improve language-model inference, \method{} constrains the latent states with product-category supervision and further uses the inferred intent categories to dynamically assemble a query-specific trie from pre-indexed category-level SID tries for retrieval decoding. Our work therefore adapts latent reasoning to a retrieval-specific setting, aiming to achieve high-performance retrieval under the latency constraints of online e-commerce systems.

\section{Preliminaries}
This section introduces the basic concepts used throughout the paper and summarizes the main notation in the proposed method. The goal is to clarify the retrieval formulation and the difference between explicit token-level reasoning and the latent intent reasoning used in \method{}.

\subsection{Basic Concepts}

\subsubsection{Generative Retrieval in E-Commerce}
Generative Retrieval (GR) reformulates retrieval as a sequence-to-sequence generation problem. Given a shopping query $q$ and a product catalog $\mathcal{D}$, the model directly generates the identifier $z_d$ of a target product item $d \in \mathcal{D}$:
\begin{equation}
    P_\theta(z_d|q) = \prod_{i=1}^{|z_d|} P_\theta(z_{d,i} | z_{d,<i}, q).
\end{equation}
The generated identifier is then mapped back to the corresponding catalog item. In practice, recent GR methods usually use Semantic IDs (SIDs), where $z_d=(s_d^1,s_d^2,\dots,s_d^m)$ is a discrete token sequence constructed by hierarchical clustering or quantization methods such as RQ-VAE~\cite{rqvae}. Compared with arbitrary item numbers, SIDs provide a more structured index because similar items may share prefix tokens. However, e-commerce GR still faces a representation gap: the query $q$ is expressed in natural language, whereas the SID $z_d$ is composed of artificial code tokens. Directly learning $P_\theta(z_d|q)$ therefore requires the model to map shopping intent into an abstract item-index space without intermediate grounding.

\subsubsection{Explicit Reasoning for Generative Retrieval}
A straightforward way to bridge this gap is to add explicit reasoning before SID generation. An explicit Chain-of-Thought (CoT) style retriever first generates a sequence of intermediate reasoning tokens $r=(r_1,\dots,r_T)$, such as inferred product categories, key attributes, or a textual description of the query intent, and then generates the target SID conditioned on these tokens:
\begin{equation}
    P_\theta(r,z_d|q)
    = \prod_{t=1}^{T} P_\theta(r_t|r_{<t},q)
      \prod_{i=1}^{|z_d|} P_\theta(z_{d,i}|z_{d,<i},r,q).
\end{equation}
In e-commerce search, this can be implemented by generating a category path before the SID, e.g., ``Electronics $\rightarrow$ Laptop Accessories $\rightarrow$ Laptop Case'' followed by the item SID. This makes the reasoning process interpretable, but it also introduces additional autoregressive decoding steps and forces the model to commit to discrete intermediate tokens before retrieval. Therefore, explicit reasoning may improve semantic grounding while sacrificing latency and robustness.

\subsubsection{Category-Guided Latent Intent Reasoning}
Latent intent reasoning aims to preserve the benefit of intermediate reasoning without explicitly generating reasoning tokens. In this paper, it does not denote a separately sampled probabilistic latent variable. Instead, it is implemented as a sequence of continuous hidden states inside the Transformer decoder before SID generation. These states are guided by the product category hierarchy so that the model can move beyond direct query-item matching and form a coarse-to-fine shopping-intent plan. Given the query encoder representations $\mathbf{H}_{\mathrm{enc}}$, the decoder performs $L$ latent reasoning steps:
\begin{equation}
    \mathbf{h}_l = \mathrm{DecoderBlock}_{\theta}(\mathbf{h}_{l-1}, \mathbf{H}_{\mathrm{enc}}), \quad l=1,\dots,L,
\end{equation}
where $\mathbf{h}_0$ is initialized by the decoder start token and $\mathbf{h}_l \in \mathbb{R}^{d_{\mathrm{model}}}$ is the $l$-th continuous reasoning state. These states form a latent intent path $\mathbf{R}_q=\{\mathbf{h}_1,\dots,\mathbf{h}_L\}$, which is used as additional context for SID decoding:
\begin{equation}
    P_\theta(z_d|q) = \prod_{i=1}^{|z_d|} P_\theta(z_{d,i} \mid z_{d,<i}, \mathbf{H}_{\mathrm{enc}}, \mathbf{R}_q).
\end{equation}
This formulation connects directly to the method section: the reasoning process is realized through Transformer hidden-state computation rather than through an explicit probability over latent variables. \method{} further supervises these hidden states with hierarchical product-category information and uses them to guide constrained SID decoding. In this way, category-guided latent intent reasoning serves as an implicit coarse-to-fine planning process while avoiding the latency and error propagation caused by explicit reasoning-token generation.

\subsection{Notation}
Table~\ref{tab:notation} summarizes the main notation used in the methodology. Bold symbols denote vectors or matrices.

\begin{table}[t]
    \centering
    \caption{Main notation used in \method{}.}
    \label{tab:notation}
    \small
    \renewcommand{\arraystretch}{1.08}
    \begin{tabularx}{\linewidth}{@{}lX@{}}
        \toprule
        \textbf{Notation} & \textbf{Description} \\
        \midrule
        $q$ & A shopping query. \\
        $\mathcal{S}_{\mathrm{train}}$, $\mathcal{B}$ & Training set and a mini-batch sampled from it. \\
        $\mathcal{D}$, $d$ & Product catalog and an item in the catalog. \\
        $z_d$ or $y$ & The Semantic ID (SID) of item $d$; $y$ is used when the SID is treated as the target generation sequence. \\
        $s_d^i$ or $s_u$ & The $i$-th or $u$-th discrete SID token/code. \\
        $m$ & Number of quantization levels used to construct an SID. \\
        $\mathbf{x}$, $\hat{\mathbf{x}}$ & Product semantic embedding and its reconstructed embedding. \\
        $E$, $D$ & Encoder and decoder in RQ-VAE. \\
        $\mathbf{z}$, $\hat{\mathbf{z}}$ & Continuous latent representation and quantized latent representation in RQ-VAE. \\
        $\mathcal{S}^{(u)}$, $\mathbf{e}_{p}^{(u)}$ & Codebook at quantization level $u$ and its $p$-th codeword. \\
        $f_{\mathrm{emb}}(\cdot)$, $V_u$ & Product text encoder used before RQ-VAE and the size of the $u$-th codebook. \\
        $\mathbf{r}_u$, $\mathbf{q}_u$ & Residual vector and selected quantized component at the $u$-th quantization level. \\
        $\lambda_{\mathrm{rq}}$ & Commitment-loss weight in RQ-VAE training. \\
        $\mathbf{H}_{\mathrm{enc}}$ & Encoder representations of query $q$. \\
        $L$, $L_d$, $l$ & Maximum latent reasoning depth, item-specific valid category depth, and a level index. \\
        $\mathbf{h}_l$, $\mathbf{R}_q$ & Hidden state at latent reasoning step $l$ and the latent intent path $\{\mathbf{h}_1,\dots,\mathbf{h}_L\}$. \\
        $\mathcal{C}_d$, $c_d^l$ & Hierarchical category path of item $d$ and its category label at level $l$. \\
        $\phi_l(\cdot)$, $\mathbf{z}_l$, $\mathbf{z}_{i,l}$ & Projection function for reasoning step $l$, the projected latent intent embedding, and its batch-indexed form for query $q_i$. \\
        $\mathbf{W}_c^l$, $\mathbf{M}(c_d^{l-1})$ & Category classifier weights at level $l$ and the hierarchy mask conditioned on the parent category. \\
        $\mathbf{v}_j$, $\mathcal{P}_{i,l}$, $\mathcal{N}_{\mathcal{B}}$ & Level-specific category prototype for category $c_j$, positive category set for query $q_i$ at level $l$, and in-batch category candidate set. \\
        $\tau$ & Temperature parameter in the multi-positive contrastive objective. \\
        $\mathcal{L}_{\mathrm{recon}}$, $\mathcal{L}_{\mathrm{codebook}}$, $\mathcal{L}_{\mathrm{commit}}$ & Reconstruction, codebook, and commitment losses in RQ-VAE training. \\
        $\mathcal{L}_{rqvae}$, $\mathcal{L}_{cls}$, $\mathcal{L}_{con}$, $\mathcal{L}_{gen}$, $\mathcal{L}_{total}$ & RQ-VAE loss, hierarchical classification loss, contrastive loss, SID generation loss, and total training loss. \\
        $\alpha$, $\beta$ & Weights for balancing hierarchical classification and contrastive losses in the total objective. \\
        $\mathcal{C}_{\text{top-}K}$, $\mathcal{D}_{\text{sub}}$ & Top-$K$ inferred categories and the candidate item subset filtered by them. \\
        $\mathcal{T}$ & Query-specific dynamic prefix trie assembled from pre-indexed category-level SID tries. \\
        $\mathbf{H}_{\text{reasoning}}$, $\mathbf{H}_{\text{full}}$ & Latent reasoning states and their concatenation with query encoder representations. \\
        $K$, $N$ & Number of inferred categories used for candidate filtering and beam/output size. \\
        $w$, $y_{<t}$ & Candidate next SID token and the previously generated SID prefix at decoding step $t$. \\
        \bottomrule
    \end{tabularx}
\end{table}

\section{Methodology}

\begin{figure}
    \centering
    \includegraphics[width=\linewidth]{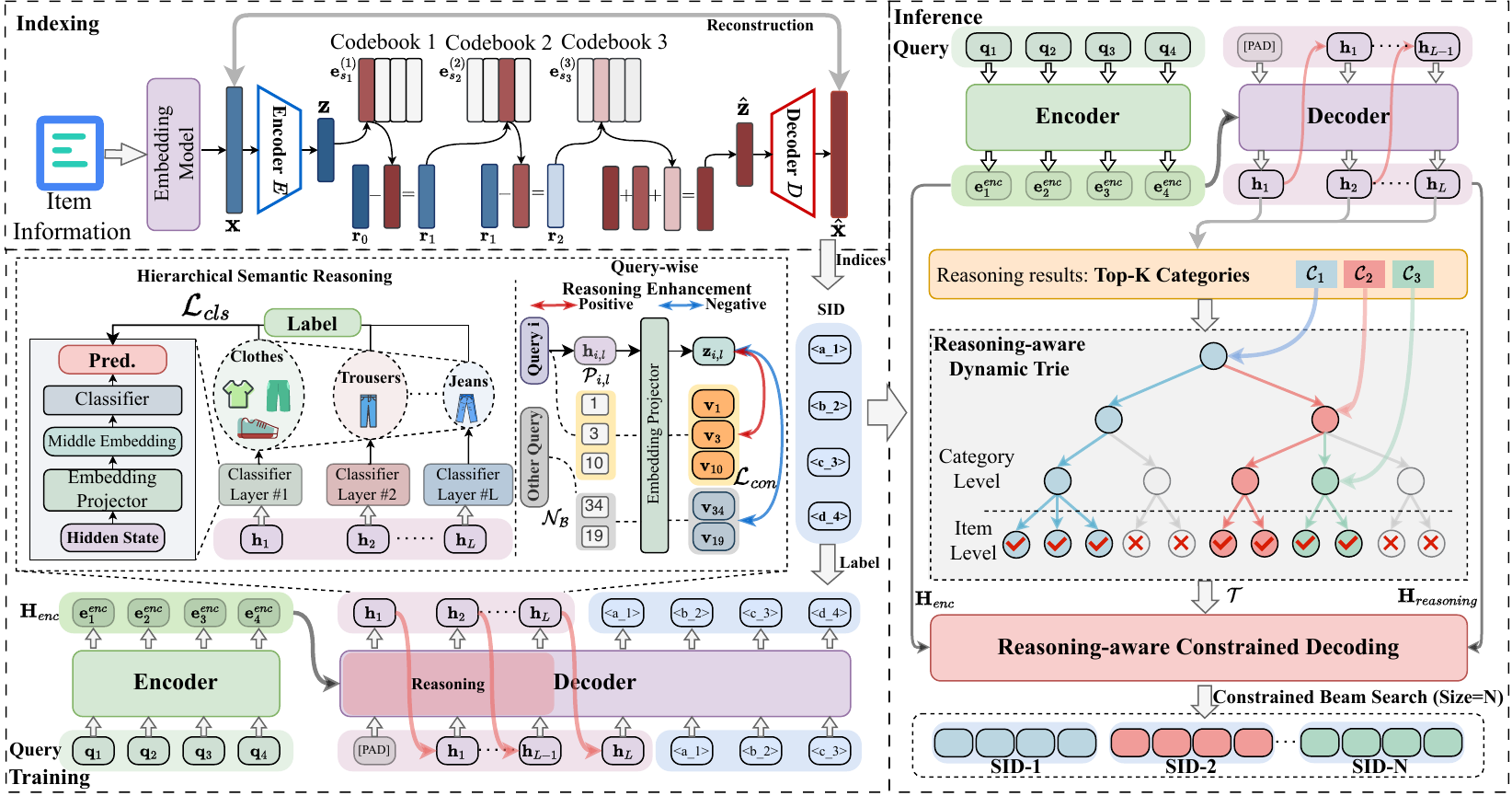}
    \caption{Overall framework of \method{}: 1) Indexing, where items are quantized into semantic IDs. 2) Training, which learns category-guided latent intent reasoning through Hierarchical Semantic Reasoning and Query-wise Reasoning Enhancement. 3) Inference, which utilizes a query-specific dynamic prefix trie assembled from pre-indexed category-level tries for precise decoding.}
    \label{fig:method}
\end{figure}

In this section, we provide a detailed description of \method{} for e-commerce search. As illustrated in Fig.~\ref{fig:method}, the method comprises three phases: indexing, training, and inference. First, the indexing phase follows a standard design where product information is encoded into embeddings and indexed via RQ-VAE~\cite{rqvae}. During the training phase, we introduce two novel learning tasks, including Hierarchical Semantic Reasoning~(HSR) and Query-wise Reasoning Enhancement~(QRE), to capture product-category intent information and guide the model in performing category-guided latent intent reasoning from coarse- to fine-grained levels. Finally, in the inference phase, we propose a Reasoning-aware Constrained Decoding~(RCD) strategy to leverage the inferred intent paths. This strategy employs constrained beam search with a query-specific dynamic prefix trie assembled from pre-indexed category-level SID tries according to latent reasoning results, enabling efficient and precise catalog retrieval.

\subsection{Indexing: Semantic ID Generation}
During the indexing phase, we employ the Residual-Quantized Variational AutoEncoder (RQ-VAE)~\cite{rqvae} to discretize high-dimensional product features into hierarchical Semantic IDs. RQ-VAE is suitable for GR because it converts a continuous product representation into a short sequence of discrete codes, while preserving a coarse-to-fine structure through multiple residual codebooks. Instead of assigning an arbitrary integer to each item, RQ-VAE decomposes the item embedding into a sum of codewords selected from different quantization levels. The selected code indices naturally form the SID used by the downstream generative retriever. The procedure consists of the following three steps.

\noindent \textbf{Embedding Phase.} Given a product item $d$ with textual fields such as title and description, we first utilize a pre-trained language model (e.g., Sentence-T5~\cite{sentence-t5}) to extract a dense semantic embedding:
\begin{equation}
    \mathbf{x} = f_{\mathrm{emb}}(d), \quad \mathbf{x} \in \mathbb{R}^{D},
\end{equation}
where $f_{\mathrm{emb}}(\cdot)$ denotes the product text encoder and $\mathbf{x}$ is the semantic representation to be indexed.

\noindent \textbf{Training Phase.} RQ-VAE first maps the dense product embedding into a lower-dimensional latent vector:
\begin{equation}
    \mathbf{z} = E(\mathbf{x}), \quad \mathbf{r}_0 = \mathbf{z},
\end{equation}
where $E$ is the RQ-VAE encoder and $\mathbf{r}_0$ is the initial residual to be quantized. This latent vector is then approximated by recursively selecting codewords from $m$ residual codebooks. Let $\mathcal{S}^{(u)}=\{\mathbf{e}_{p}^{(u)}\}_{p=1}^{V_u}$ denote the codebook at quantization level $u$, where $V_u$ is the number of codewords in that codebook. At each level, RQ-VAE selects the nearest codeword to the current residual:
\begin{equation}
    s_u = \arg\min_{p \in \{1,\dots,V_u\}} \|\mathbf{r}_u - \mathbf{e}_{p}^{(u)}\|_2,
\end{equation}
and uses the selected codeword as the level-$u$ quantized component:
\begin{equation}
    \mathbf{q}_u = \mathbf{e}_{s_u}^{(u)}.
\end{equation}
The residual is then updated by subtracting the selected component:
\begin{equation}
    \mathbf{r}_{u+1} = \mathbf{r}_u - \mathbf{q}_u.
\end{equation}
This recursive process makes early codebooks capture larger semantic factors, while later codebooks focus on the remaining fine-grained residual information. After all $m$ levels, the quantized latent representation is obtained by summing all selected codewords:
\begin{equation}
    \hat{\mathbf{z}} = \sum_{u=0}^{m-1} \mathbf{q}_u
    = \sum_{u=0}^{m-1} \mathbf{e}_{s_u}^{(u)}.
\end{equation}
Thus, the discrete code tuple $(s_0,\dots,s_{m-1})$ is a compact symbolic representation of item $d$, while $\hat{\mathbf{z}}$ is its continuous reconstruction in the latent space. We use $u$ for RQ-VAE quantization levels to distinguish them from the category-reasoning levels used later in \method{} training.

To ensure that these discrete codes retain the essential information of the original product embedding, a decoder $D$ reconstructs the input embedding from the quantized latent representation:
\begin{equation}
    \hat{\mathbf{x}} = D(\hat{\mathbf{z}}).
\end{equation}
The RQ-VAE objective can be understood as three components. The first component is the reconstruction loss, which preserves the semantic content of the product embedding:
\begin{equation}
    \mathcal{L}_{\mathrm{recon}} = \|\mathbf{x} - \hat{\mathbf{x}}\|_2^2.
\end{equation}
The second component is the codebook loss, which moves the selected codewords toward the residual vectors assigned to them:
\begin{equation}
    \mathcal{L}_{\mathrm{codebook}}
    = \sum_{u=0}^{m-1}
    \|\mathrm{sg}[\mathbf{r}_u] - \mathbf{e}_{s_u}^{(u)}\|_2^2.
\end{equation}
The third component is the commitment loss, which prevents the encoder output from drifting away from the selected codewords:
\begin{equation}
    \mathcal{L}_{\mathrm{commit}}
    = \sum_{u=0}^{m-1}
    \|\mathbf{r}_u - \mathrm{sg}[\mathbf{e}_{s_u}^{(u)}]\|_2^2.
\end{equation}
Here, $\mathrm{sg}[\cdot]$ denotes the stop-gradient operator. It separates the update of the codebook vectors from the update of the encoder representation during backpropagation. The final RQ-VAE training objective is:

\begin{equation}
    \mathcal{L}_{rqvae}
    = \mathcal{L}_{\mathrm{recon}}
    + \mathcal{L}_{\mathrm{codebook}}
    + \lambda_{\mathrm{rq}} \mathcal{L}_{\mathrm{commit}},
\end{equation}
where $\lambda_{\mathrm{rq}}$ controls how strongly the encoder is encouraged to commit to the selected residual codewords. By optimizing this objective, RQ-VAE learns codebooks that reconstruct item semantics while producing compact discrete identifiers.

\noindent \textbf{Generation Phase.} 
Upon convergence of the RQ-VAE training, we generate a Semantic ID (SID) for each item by running the encoder and residual quantizer once and concatenating the selected code indices:
\begin{equation}
    \mathrm{SID}(d) = (s_0, s_1, \dots, s_{m-1}).
\end{equation}
For example, an item may be represented as ``<a\_1><b\_2><c\_3><d\_4>''. To address potential collisions where multiple items obtain the same SID, we adopt the Sinkhorn algorithm~\cite{cuturi2013sinkhorn} to re-assign unique identifiers. These final SIDs serve as the target sequences for \method{} training and the valid paths used in constrained decoding.

\subsection{Training: Category-Guided Latent Intent Reasoning}
\label{sec:training}

To bridge the semantic gap between shopping queries and discrete item SIDs, we conceptualize latent intent reasoning as a category-guided ``mental planning'' phase. Rather than mapping $q \rightarrow \text{SID}$ directly, the model internalizes hierarchical shopping intents to recalibrate its hidden states. This aligns the decoder's trajectory with the target catalog subspace prior to generating concrete SIDs.

This pre-generation placement is critical. Once an autoregressive GR model emits the first SID token, the decoder has already committed to a branch in the discrete index space. Since SID tokens are artificial codes rather than natural-language semantic units, an erroneous early prefix may lead beam search to concentrate on an irrelevant catalog region, making later tokens or post-hoc reranking difficult to recover the correct item. Performing reasoning before the first SID token allows the model to calibrate its decoder trajectory while it is still operating in the query-conditioned continuous representation space.

A straightforward alternative is to generate an explicit category path or textual rationale before the SID. However, such explicit reasoning introduces additional autoregressive steps and forces the model to commit to discrete intermediate labels, which is undesirable for low-latency e-commerce search and for ambiguous multi-intent queries. \method{} instead inserts a fixed number of category-guided latent reasoning steps before SID decoding. These continuous states can encode coarse department-level intent, fine-grained product-family constraints, and multiple plausible category directions without exposing them as output tokens. They are then used both as supervised latent states during training and as intent signals for reasoning-aware constrained decoding during inference.

\subsubsection{Continuous Latent Intent Reasoning Process}

Previous GR models typically decode target identifiers immediately following query comprehension. However, shopping intents often require multi-step reasoning that transitions from broad needs to granular category constraints before accurately pinpointing a specific catalog item. Directly generating discrete SID tokens to model this inferential process might lead to error propagation and cumulative bias. To circumvent these issues, we perform intent reasoning within a continuous latent space, which enables the model to preserve rich semantic information and prevents premature collapse into discrete representations.

Here, we extend the T5 decoder to perform $L$ steps of latent intent reasoning. Given the encoder representations $\mathbf{H}_{\mathrm{enc}}$, the decoder initializes with a start token $\mathbf{h}_0 = \text{[PAD]}$. For each reasoning step $l \in \{1, \dots, L\}$, the evolution of the hidden state is defined as:
\begin{equation}
    \mathbf{h}_l = \mathrm{DecoderBlock}(\mathbf{h}_{l-1}, \mathbf{H}_{\mathrm{enc}}),
\end{equation}
where $\mathbf{h}_l \in \mathbb{R}^{d_{\mathrm{model}}}$ is the hidden state at step $l$. These states form a continuous latent intent path that precedes the final item SID generation.

\subsubsection{Hierarchical Semantic Reasoning}
Although the reasoning hidden states $\mathbf{h}_l$ possess the capacity to encode information, unconstrained recurrence does not inherently guarantee specific intent paths. Let $\mathcal{C}_d = \{c_d^1, c_d^2, \dots, c_d^{L_d}\}$ denote the hierarchical category path for item $d$, where $L_d$ is the valid category depth of this item and $L$ is the maximum reasoning depth used by the model. We leverage this path to facilitate coarse-to-fine intent learning. To enforce that the $l$-th latent reasoning step aligns with the semantics of the $l$-th hierarchical category level, we introduce a hierarchical semantic alignment objective.

We first project the decoder hidden state into a specialized semantic space using an embedding projector $\phi_l(\cdot)$ for each reasoning step, which serves to bridge the gap between the transformation of the reasoning process and the categorical distribution required for classification:
\begin{equation}
    \mathbf{z}_l = \phi_l(\mathbf{h}_l),
\end{equation}
where $\mathbf{z}_l$ is the projected embedding. We then define a learnable weight matrix $\mathbf{W}_c^l$ for the $l$-th level categories. To strictly enforce hierarchical dependencies and prevent logical inconsistencies (e.g., predicting a child class unrelated to the parent), we incorporate a hierarchy constraint mask $\mathbf{M}$. The probability of the item category at level $l$ is modeled as:
\begin{equation}
    P(c_d^l | \mathbf{z}_l, c_d^{l-1}) = \text{Softmax}(\mathbf{W}_c^l \mathbf{z}_l + \mathbf{M}(c_d^{l-1})),
\end{equation}
where $\mathbf{M}(c_d^{l-1})$ sets the logits of invalid sub-categories to $-\infty$ based on the parent category $c_d^{l-1}$ of the item. For the first level, $c_d^0$ denotes the virtual root of the hierarchy.

Finally, considering that the depth of the hierarchical categories may vary across different items (i.e., not all items reach the maximum depth $L$), we utilize a masked Cross-Entropy loss. This ensures that supervision is only applied to valid levels within a batch:
\begin{equation}
    \mathcal{L}_{\text{cls}} = \sum_{l=1}^{L} \mathbb{I}(l \le L_d) \cdot [-\log P(c_d^l | \mathbf{z}_l, c_d^{l-1})],
\end{equation}
where $c_d^l$ denotes the ground-truth category at layer $l$, and $\mathbb{I}(\cdot)$ is the indicator function that ignores the loss for levels beyond the given item's actual depth $L_d$. Through this module, the model initiates a progressive reasoning process, traversing from coarse-grained categories to fine-grained taxonomies, thereby imposing structured latent constraints via the attention mechanism.

\subsubsection{Query-wise Reasoning Enhancement}
In e-commerce scenarios, a single search query may correspond to multiple valid product categories (i.e., multi-positive intents), which may share a common parent category but differ at fine-grained levels. To capture the commonalities among diverse items under the same query and robustly align the latent intent states with all relevant items, we propose the Query-wise Reasoning Enhancement (QRE) mechanism. This mechanism is designed to impose geometric regularization on the latent hidden states.

Formally, we treat the row vectors of the classification weight matrix $\mathbf{W}_c^l$ in HSR module as the category embeddings (anchors). Let $\mathbf{v}_j$ denote the weight vector corresponding to category $c_j$. For the $i$-th query $q_i$ in a batch $\mathcal{B}$, at reasoning step $l$, let $\mathcal{P}_{i,l}$ denote the set of ground-truth positive category indices at the $l$-th layer. We compute the cosine similarity between the projected embedding $\mathbf{z}_{i,l}$ and a category prototype $\mathbf{v}_j$:
\begin{equation}
    s(q_i, c_j) = \frac{\mathbf{z}_{i,l}^\top \mathbf{v}_j}{\|\mathbf{z}_{i,l}\| \|\mathbf{v}_j\| \cdot \tau},
\end{equation}
where $\tau$ is a temperature hyperparameter controlling the concentration of the distribution.

Since a query may align with multiple categories, we employ a Multi-Positive InfoNCE loss. This objective maximizes the similarity between the query and all its positive prototypes while simultaneously pushing away the prototypes of irrelevant categories. The loss for query $q_i$ at layer $l$ is defined as:
\begin{equation}
    \ell_{\text{cl}}^{(i,l)} = - \frac{1}{|\mathcal{P}_{i,l}|} \sum_{p \in \mathcal{P}_{i,l}} \log \frac{\exp(s(q_i, c_p))}{\sum_{c_j \in \mathcal{N}_{\mathcal{B}}} \exp(s(q_i, c_j))},
\end{equation}
where $\mathcal{N}_{\mathcal{B}}$ represents the set of all category indices appearing in the current batch (serving as the candidate pool). 

The total contrastive loss accumulates the reasoning signals across all valid depth levels for the batch:
\begin{equation}
    \mathcal{L}_{\text{con}} = \sum_{i \in \mathcal{B}} \sum_{l=1}^L \mathbb{I}(l \le L_{d_i}) \cdot \ell_{\text{cl}}^{(i,l)}.
\end{equation}

By optimizing this objective, the model learns to cluster the query latent states tightly around the geometric centers of their associated category prototypes, thereby enhancing the discriminative power of the hierarchical reasoning process.

\subsubsection{Joint Optimization Objective}
Upon completion of the category-guided latent intent reasoning phase, the final hidden state $\mathbf{h}_L$ encapsulates the refined intent information. This state serves as the context for the decoder to generate the target item identifier sequence $y$ in an auto-regressive manner. The generative loss is defined as:
\begin{equation}
    \mathcal{L}_{\text{gen}} = -\log P(y \mid \mathbf{h}_L).
\end{equation}

The overall training objective integrates the generative task with the explicit reasoning constraints:
\begin{equation}
    \mathcal{L}_{\text{total}} = \mathcal{L}_{\text{gen}} + \alpha \mathcal{L}_{\text{cls}} + \beta \mathcal{L}_{\text{con}},
\end{equation}
where $\alpha$ and $\beta$ are hyperparameters that balance the contribution of the hierarchical semantic supervision and the contrastive geometric structuring, respectively.

Algorithm~\ref{alg:calir_training} summarizes the training procedure of \method{}.
\begin{algorithm}[htbp]
    \caption{Training Procedure of \method{}}
    \label{alg:calir_training}
    \begin{algorithmic}[1]
        \REQUIRE Training set $\mathcal{S}_{\mathrm{train}}$, where each example contains query $q_i$, target SID $y_i$, category path $\mathcal{C}_{d_i}$, and multi-positive category sets $\{\mathcal{P}_{i,l}\}_{l=1}^{L_{d_i}}$.
        \REQUIRE Model parameters $\theta$, Latent steps $L$, Hyperparameters $\alpha, \beta, \tau$.
        \WHILE{not converged}
            \FOR{each batch $\mathcal{B}$ sampled from $\mathcal{S}_{\mathrm{train}}$}
                \STATE \textbf{// 1. Encode Query}
                \STATE $\mathbf{H}_{\mathrm{enc}} \leftarrow \text{Encoder}(q)$
                \STATE Initialize latent state input $\mathbf{h}_0$ using decoder input embeddings
                \STATE Initialize losses: $\mathcal{L}_{cls} \leftarrow 0, \mathcal{L}_{con} \leftarrow 0$
                \STATE \textbf{// 2. Latent Reasoning Loop (Coarse-to-Fine)}
                \FOR{$l = 1$ to $L$}
                    \STATE $\mathbf{h}_l \leftarrow \text{DecoderBlock}(\mathbf{h}_{l-1}, \mathbf{H}_{\mathrm{enc}})$
                    \STATE Project latent state: $\mathbf{z}_l \leftarrow \text{Projector}_l(\mathbf{h}_l)$
                    \STATE Get class prototypes: $\mathbf{W}_c^l \leftarrow \text{Classifier}_l.\text{weight}$
                    \STATE Compute logits: $logits_l \leftarrow \text{Classifier}_l(\mathbf{z}_l)$
                    \STATE $\mathcal{L}_{cls} \leftarrow \mathcal{L}_{cls} + \text{CrossEntropy}(logits_l, c_{d_i}^{l})$
                    \IF{multi-positive category set $\mathcal{P}_{i,l}$ exists}
                        \STATE Retrieve positive category embeddings: $\mathbf{E}_{pos} \leftarrow \text{Lookup}(\mathcal{P}_{i,l}, \mathbf{W}_c^l)$
                        \STATE $\mathcal{L}_{con} \leftarrow \mathcal{L}_{con} + \text{MultiPosInfoNCE}(\mathbf{z}_l, \mathbf{E}_{pos}, \tau)$
                    \ENDIF
                \ENDFOR
                \STATE \textbf{// 3. Generative Retrieval (SID Generation)}
                \STATE $\mathbf{H}_{latent} \leftarrow \mathbf{h}_L$
                \STATE Generate token sequence predictions $\hat{y}$ conditioned on $\mathbf{H}_{latent}$
                \STATE $\mathcal{L}_{gen} \leftarrow \text{CrossEntropy}(\hat{y}, y)$
                \STATE \textbf{// 4. Optimization}
                \STATE $\mathcal{L}_{total} \leftarrow \mathcal{L}_{gen} + \alpha \cdot \mathcal{L}_{cls} + \beta \cdot \mathcal{L}_{con}$
                \STATE Update $\theta \leftarrow \theta - \eta \nabla_\theta \mathcal{L}_{total}$
            \ENDFOR
        \ENDWHILE
    \end{algorithmic}
\end{algorithm}

\subsection{Inference: Reasoning-aware Constrained Decoding}

During inference, we transform the continuous latent intent reasoning signals into discrete item identifiers through a synchronized two-stage mechanism. This process combines structural pruning with semantic guidance: 1) \textbf{Reasoning-aware Dynamic Trie Construction}, which narrows the search space by assembling a query-specific trie constraint from pre-indexed category-level SID tries according to predicted intent categories. 2) \textbf{Constrained Beam Search}, which generates the final item SIDs by integrating the information from latent reasoning within the dynamic trie constraint.

Algorithm~\ref{alg:calir_inference} summarizes the inference procedure of \method{}.
\begin{algorithm}[!htbp]
    \caption{Inference Procedure of \method{}}
    \label{alg:calir_inference}
    \begin{algorithmic}[1]
        \REQUIRE Query $q$, Model parameters $\theta$, pre-indexed category-level SID tries $\{\mathcal{T}_c\}_{c\in\mathcal{C}}$.
        \REQUIRE Beam size $N$, Top-$K$ reasoning categories.
        \STATE $\mathbf{H}_{\mathrm{enc}} \leftarrow \text{Encoder}(q)$
        \STATE Initialize latent state $\mathbf{h}_0$
        \FOR{$l = 1$ to $L$}
            \STATE $\mathbf{h}_l \leftarrow \text{DecoderBlock}(\mathbf{h}_{l-1}, \mathbf{H}_{\mathrm{enc}})$
        \ENDFOR
        \STATE Project final state: $\mathbf{z}_L \leftarrow \text{Projector}_L(\mathbf{h}_L)$
        \STATE Predict category distribution: $P(c|\mathbf{z}_L) \leftarrow \text{Softmax}(\text{Classifier}_L(\mathbf{z}_L))$
        \STATE Get Top-$K$ categories: $\mathcal{C}_{\text{top-}K} \leftarrow \text{TopK}(P(c|\mathbf{z}_L), K)$
        \STATE Assemble dynamic trie constraint $\mathcal{T} \leftarrow \mathrm{Assemble}(\{\mathcal{T}_c \mid c \in \mathcal{C}_{\text{top-}K}\})$
        \STATE Generate candidates $y$ using Beam Search constrained by $\mathcal{T}$
        \RETURN Top-$N$ generated SIDs
    \end{algorithmic}
\end{algorithm}

\subsubsection{Reasoning-aware Dynamic Trie Construction}

To efficiently navigate the large item space, we construct a Reasoning-aware Dynamic Prefix Trie for each query. The trie is dynamic because its active branches depend on the categories predicted by latent reasoning for the current query. In implementation, however, we pre-index category-level SID prefix tries before online serving. For each category $c$, the offline index stores a trie $\mathcal{T}_c$ built from the SIDs of products that belong to this category. During inference, \method{} assembles the query-specific trie by activating the top-$K$ pre-indexed category tries and taking their logical union, rather than by inserting all candidate SIDs online.

Formally, based on the probability distribution from the final reasoning step, we identify the top-$K$ semantic categories, denoted as $\mathcal{C}_{\text{top-}K}$. A larger $K$ is employed here to preserve intent diversity and recall. The dynamic trie constraint is then assembled from the corresponding pre-indexed tries:
\begin{equation}
    \mathcal{T}
    = \mathrm{Assemble}\big(\{\mathcal{T}_c \mid c \in \mathcal{C}_{\text{top-}K}\}\big),
\end{equation}
where $\mathrm{Assemble}(\cdot)$ denotes a lightweight lookup and logical union over pre-indexed trie roots. Thus, ``construction'' refers to forming the query-specific active constraint, while the expensive construction of category-level trie structures is completed offline. This dynamic trie defines the valid transition space for the subsequent decoding phase, ensuring that any generated path corresponds to an item under the inferred intent categories.

\subsubsection{Constrained Beam Search with Latent Intent Reasoning}

With the search space explicitly pruned by the dynamic prefix trie $\mathcal{T}$, the decoder must now generate the precise SIDs. Since the latent reasoning phase operates in a continuous space, we introduce a reasoning-guided decoding strategy that bridges continuous reasoning states with discrete token generation.

Specifically, we utilize the sequence of hidden states $\mathbf{H}_{\text{reasoning}} = \{\mathbf{h}_1, \dots, \mathbf{h}_L\}$ obtained from the reasoning phase. These are concatenated with the query encoder outputs $\mathbf{H}_{\mathrm{enc}}$ to integrate the information from latent reasoning, as shown below,
\begin{equation}
    \mathbf{H}_{\text{full}} = [\mathbf{H}_{\mathrm{enc}}; \mathbf{H}_{\text{reasoning}}].
\end{equation}
In implementation, this reasoning stage is performed once for the whole batch before beam expansion. The resulting batch-level latent states are then concatenated with the corresponding query encoder states and used as the shared context for the subsequent batched beam search. Therefore, \method{} adds a fixed number of continuous reasoning steps, but does not branch over latent states during beam search.

The decoder performs a constrained beam search over the prefix trie $\mathcal{T}$. At each decoding step $t$, the probability of a token $w$ is conditioned on the fused hidden state $\mathbf{H}_{\text{full}}$, but strictly masked by the valid paths in $\mathcal{T}$:
\begin{equation}
    P(w | y_{<t}) = 
    \begin{cases} 
    \text{Softmax}(\mathbf{W}_o \cdot \text{Dec}(w, y_{<t}, \mathbf{H}_{\text{full}})), & \text{if } (y_{<t}, w) \in \mathcal{T}, \\
    0, & \text{otherwise,}
    \end{cases}
\end{equation}
where $Dec$ represents the decoder of \method{}.

By conditioning the generation on $\mathbf{H}_{\text{full}}$ while enforcing the structural constraints of $\mathcal{T}$, the retrieval process effectively incorporates both the hierarchical reasoning logic and the validity of the target items.
Finally, we generate $N$ SIDs as the retrieved results, where $N$ represents the size of beam search. These SIDs are then mapped back to their corresponding original items to produce the final retrieval set.

\subsection{Complexity Analysis}

We analyze the time complexity of \method{} to clarify the additional cost introduced by latent reasoning. The RQ-VAE indexing stage is an offline preprocessing step and follows the standard residual quantization design used in prior GR models such as TIGER. For a batch of item embeddings, the naive nearest-codeword search across $m$ residual codebooks costs $\mathcal{O}(\sum_{u=0}^{m-1} V_u d_z)$ per item, where $V_u$ is the size of the $u$-th codebook and $d_z$ is the latent dimensionality. This cost is paid only when constructing item SIDs and is not on the online inference path.

For online retrieval, let $B$ be the batch size, $n$ be the query length, $m$ be the SID length, $L$ be the number of latent reasoning steps, $N$ be the beam size, $K$ be the number of selected intent categories, and $R$ be the number of explicit reasoning tokens in a CoT-style retriever. Let $C_{\mathrm{enc}}(n)$ denote the cost of encoding a query of length $n$, and let $C_{\mathrm{dec}}(s)$ denote the cost of one decoder step attending to a context of length $s$. We also use $\bar{a}_{\mathrm{all}}$ and $\bar{a}_{\mathrm{sub}}$ to denote the average number of valid next-token transitions in the global SID trie and in the query-specific dynamic trie, respectively. Since the dynamic trie covers only the top-$K$ inferred categories, normally $\bar{a}_{\mathrm{sub}} \le \bar{a}_{\mathrm{all}}$. The comparison in Table~\ref{tab:complexity_comparison} omits constant factors and the number of Transformer layers. For compactness, define $E=B C_{\mathrm{enc}}(n)$, $G_s=BNm C_{\mathrm{dec}}(s)$, $R_{\mathrm{tok}}=BR C_{\mathrm{dec}}(n)$, $R_{\mathrm{lat}}=BL C_{\mathrm{dec}}(n)$, $A_{\mathrm{all}}=BNm\bar{a}_{\mathrm{all}}$, and $A_{\mathrm{sub}}=BNm\bar{a}_{\mathrm{sub}}$. We use $C_{\mathrm{dyn}}$ to denote the online cost of assembling the query-specific dynamic trie from pre-indexed category tries.

\begin{table}[!htbp]
    \centering
    \caption{Online inference complexity comparison.}
    \label{tab:complexity_comparison}
    \scriptsize
    \renewcommand{\arraystretch}{1.12}
    \begin{tabularx}{\linewidth}{@{}p{2.1cm}X X@{}}
        \toprule
        \textbf{Model} & \textbf{Main online procedure} & \textbf{Time complexity per batch} \\
        \midrule
        TIGER & Encode query and directly run constrained beam search over the global SID trie. &
        $\mathcal{O}(E + G_n + A_{\mathrm{all}})$ \\
        Explicit CoT + GR & Autoregressively generate explicit reasoning tokens before SID decoding. &
        $\mathcal{O}(E + R_{\mathrm{tok}} + G_{n+R} + A_{\mathrm{all}})$ \\
        \method{} & Batch latent reasoning, dynamic trie construction from pre-indexed category tries, and batched beam search with fused latent context. &
        $\mathcal{O}(E + R_{\mathrm{lat}} + C_{\mathrm{dyn}} + G_{n+L} + A_{\mathrm{sub}})$ \\
        \bottomrule
    \end{tabularx}
\end{table}

Here, $C_{\mathrm{dyn}}$ denotes the lightweight online assembly of the query-specific dynamic trie. Since category-level SID tries are pre-indexed offline, this step mainly activates and unions the top-$K$ trie roots for each query, giving $C_{\mathrm{dyn}}=\mathcal{O}(BK)$ under direct trie-root lookup. It does not incur the online $\mathcal{O}(B|\mathcal{D}_{\mathrm{sub}}|m)$ cost that would be required if the trie were rebuilt by inserting all filtered candidate SIDs. The valid-transition work after assembly is already included in $A_{\mathrm{sub}}$. The key difference from explicit CoT is that \method{} computes $L$ latent states once at the batch level and then reuses them as context for all beams. Thus, the latent reasoning cost is linear in $BL$ and is not multiplied by the beam size $N$ or the SID length $m$. Compared with TIGER, \method{} introduces this fixed latent-reasoning term and lightweight dynamic-trie assembly, while the query-specific trie can reduce the valid transition space during SID beam search.

\FloatBarrier

\section{Experiments}
In this section, we analyze our model's effectiveness and address the following research questions:
\begin{itemize}
    \item \textbf{RQ1:} How does \method{} compare to state-of-the-art baselines?
    \item \textbf{RQ2:} How do different modules impact \method{}'s performance?
    \item \textbf{RQ3:} How do key hyperparameters influence \method{}'s retrieval performance?
    \item \textbf{RQ4:} Does the \method{} method demonstrate superior performance compared to explicit CoT?
    \item \textbf{RQ5:} Does \method{} remain effective when paired with different e-commerce item SID construction methods?
    \item \textbf{RQ6:} How does the step length of category-guided latent intent reasoning influence the quality of retrieval results?
    \item \textbf{RQ7:} Does \method{} effectively capture category-level shopping intent beyond direct query-item matching?
    \item \textbf{RQ8:} Can \method{} transfer to general-domain retrieval when predefined product taxonomies are unavailable?
    \item \textbf{RQ9:} Does \method{} remain effective under a different generative backbone?
    \item \textbf{RQ10:} How does \method{} perform in terms of inference efficiency and latency compared to baseline models?
\end{itemize}

\subsection{Experimental Setup}
\subsubsection{Datasets.}
\label{ap_dataset}
We use ESCI~\cite{esci}, a multilingual e-commerce search benchmark released by Amazon, for query-product semantic matching. It contains authentic search queries, product catalogs, and E-S-C-I relevance labels, and it includes complex negation and product-attribute constraints. These properties cause low term overlap between queries and products, making the dataset a challenging benchmark for e-commerce search. After filtering a subset of items based on category information, the resulting dataset scale is presented in Table~\ref{tab:dataset_stats}.

\begin{table}[htbp]
\centering
\setlength\tabcolsep{1pt}
\caption{Statistics of three datasets.}
\begin{tabular}{lccccc}
\toprule
  && \multicolumn{2}{c}{Train} & \multicolumn{2}{c}{Test} \\
\cmidrule(lr){3-4} \cmidrule(lr){5-6}
Dataset & \#Products & \#Queries & \#Q-P Pairs & \#Queries & \#Q-P Pairs \\
\midrule
ESCI-us & 288,372 & 20,109 & 292,354 & 6,137 & 29,971 \\
ESCI-es & 101,957 & 5,421 & 108,587 & 1,877 & 14,554 \\
ESCI-jp & 119,052 & 6,543 & 127,788 & 2,163 & 15,861 \\
\bottomrule
\end{tabular}
\label{tab:dataset_stats}
\end{table}

\subsubsection{Baselines.}
\label{ap_baselines}
We compare \method{} with three categories of baselines, including sparse retrieval, dense retrieval, and generative retrieval (GR). The details are listed as follows:
\begin{itemize}
    \item \textit{BM25}~\cite{bm25}: a sparse retrieval method based on statistical term-frequency matching, which remains effective when exact product attributes are important.
    \item \textit{DPR}~\cite{dpr}: a dense retrieval method that maps queries and products into dense vector spaces for semantic matching beyond surface-form overlap.
    \item \textit{Sentence-T5}~\cite{sentence-t5}: a sentence encoder built from pre-trained text-to-text models, which provides dense semantic representations for product matching.
    \item \textit{MPNet}~\cite{song2020mpnet}: a pre-trained language model that combines masked language modeling and permuted language modeling to learn contextual text representations for dense retrieval.
    \item \textit{BGE-M3}~\cite{chen2024m3}: a versatile text embedding model that unifies dense, sparse, and multi-vector retrieval functionalities, while supporting multilingual and long-text product encoding.
    \item $\text{DSI}_{\text{naive}}$~\cite{dsi}: a generative retrieval method that maps a natural language query to an atomic item identifier.
    \item $\text{DSI}_{\text{semantic}}$~\cite{dsi}: a generative retrieval method that constructs structured identifiers to preserve semantic information in the identifier space.
    \item \textit{TIGER}~\cite{rqvae}: a generative retrieval framework that constructs item Semantic IDs by quantizing item content embeddings with RQ-VAE, and trains a Transformer-based sequence-to-sequence model to autoregressively generate target item identifiers.
    \item \textit{Hi-Gen}~\cite{hi-gen}: a hierarchical generative retrieval method that employs clustering-based structures to preserve hierarchical relationships among items.
    \item \textit{LTRGR}~\cite{ltrgr}: a generative retrieval method that incorporates ranking signals to optimize the retrieval model.
    \item \textit{RIPOR}~\cite{ripor}: a scalable generative retrieval framework designed to improve the effectiveness and efficiency of GR.
    \item \textit{MERGE}~\cite{merge}: a generative retrieval method that introduces multi-level relevance alignment during SID learning to enhance identifier representation.
    \item \textit{CAT-ID}$^2$~\cite{catid}: an e-commerce GR method that embeds multi-level category information into identifier learning, encouraging identifiers to reflect product category distributions.
\end{itemize}

\subsubsection{Evaluation Metrics.}
Following the common evaluation protocol in recommendation and product retrieval, we treat each query as a recommendation request and evaluate whether the model can rank its relevant products at the top positions. Let $\mathcal{Q}$ denote the test query set, $\mathcal{G}_q$ denote the ground-truth relevant product set for query $q$, and $\pi_q$ denote the ranked product list returned by the model. The top-$k$ results are denoted as $\pi_q^{1:k}$.

We use \textbf{Recall@$k$/R@$k$} ($k \in \{5, 10, 100\}$) to measure retrieval coverage:
\begin{equation}
    \mathrm{Recall@}k = \frac{1}{|\mathcal{Q}|}\sum_{q \in \mathcal{Q}}
    \frac{|\pi_q^{1:k} \cap \mathcal{G}_q|}{|\mathcal{G}_q|}.
\end{equation}
This metric is suitable for e-commerce retrieval because a shopping query can correspond to multiple relevant products. Therefore, the metric measures how many relevant products are covered by the top-$k$ retrieved results, rather than only checking a single target item.

We use \textbf{NDCG@$k$/N@$k$} ($k \in \{10, 100\}$) to evaluate position-aware ranking quality. For a product at rank $j$, we define $\mathrm{rel}_{q,j}=1$ if $\pi_q[j]\in\mathcal{G}_q$, and $\mathrm{rel}_{q,j}=0$ otherwise. The DCG and ideal DCG are computed as:
\begin{equation}
    \mathrm{DCG@}k(q) = \sum_{j=1}^{k}\frac{\mathrm{rel}_{q,j}}{\log_2(j+1)},
\end{equation}
\begin{equation}
    \mathrm{IDCG@}k(q) = \sum_{j=1}^{\min(k,|\mathcal{G}_q|)}
    \frac{1}{\log_2(j+1)}.
\end{equation}
Then NDCG@$k$ is defined as:
\begin{equation}
    \mathrm{NDCG@}k = \frac{1}{|\mathcal{Q}|}\sum_{q \in \mathcal{Q}}
    \frac{\mathrm{DCG@}k(q)}{\mathrm{IDCG@}k(q)}.
\end{equation}
NDCG gives higher scores when relevant products appear at earlier ranks, which matches the practical requirement of product search and recommendation. Unless otherwise specified, metric values in the ESCI experiment tables are reported as percentages.

\subsubsection{Implementation Details.}
For the main ESCI experiments, we employ a pre-trained T5-base model as the backbone for English GR and a pre-trained multilingual T5-base model for other languages. For product representation, we employ BERT and mBERT to encode item information into embeddings for English and other languages, respectively. For the codebook configuration, we define the item SID length as 4, which requires four codebooks, each with a size of 256. The RQ-VAE is trained for 300 epochs using a batch size of 2048 and the AdamW optimizer. For GR training, we use a per-device batch size of 512, a learning rate of $5\times 10^{-4}$, and train for 300 epochs on NVIDIA A100 80GB GPUs. Regarding the training of the default GR model, we set the hyperparameters $\alpha$ and $\beta$ to 0.1 across the ESCI datasets. For the Reasoning-aware Dynamic Trie, the number of top-$K$ reasoning categories is fixed at $K=3$; category-level tries are pre-indexed offline, and the query-specific trie is assembled online by activating the top-$K$ tries. We repeat the main retrieval experiments multiple times with different random seeds and conduct paired t-tests. In the main result table, $^*$ denotes statistically significant improvement over the strongest baseline with $p<0.05$.

For baseline reproduction, we use official source code when it is available and re-implement methods without public repositories. For the DSI models, $\text{DSI}_{\text{naive}}$ assigns IDs sequentially based on product catalog order, while $\text{DSI}_{\text{semantic}}$ constructs SIDs via a $20 \times 20 \times 20$ hierarchical $K$-means algorithm with a suffix added to resolve ID collisions. For TIGER~\footnote{https://github.com/HonghuiBao2000/LETTER}, we use the RQ-VAE method from LETTER~\cite{letter} to construct SIDs. For LTRGR~\footnote{https://github.com/liyongqi67/LTRGR} and RIPOR~\footnote{https://github.com/HansiZeng/RIPOR}, we follow the official execution logic. For Hi-GEN, due to the lack of official code, we reconstruct its hierarchy by using item category information for the first two layers, followed by two layers generated through hierarchical $K$-means and a final collision-resolution layer. For MERGE~\footnote{https://github.com/zhangfw123/MERGE}, we use the official code for SID construction. The SIDs for $\text{CAT-ID}^2$ are obtained through direct correspondence with the authors and filtered to retain only the items present in our dataset for training.

\subsection{Main Retrieval Performance (RQ1)}
\begin{table}[t]
  \centering
  \setlength\tabcolsep{2.5pt}
  \caption{Experimental Results on ESCI-us, ESCI-es, and ESCI-jp datasets. Baselines are categorized into sparse, dense, and generative retrieval models. The best results are highlighted in \textbf{bold}, while the second-best results are \underline{underlined} in all groups. The symbol $^*$ after a \method{} score indicates statistically significant improvement over the strongest baseline for the corresponding metric under a paired t-test ($p<0.05$).}
  \label{tab:performance_comparison}
  \resizebox{\textwidth}{!}{
  \begin{tabular}{lccccccccccccccc}
    \toprule
    \multirow{2}{*}{\textbf{Model}} & \multicolumn{5}{c}{\textbf{ESCI-us}} & \multicolumn{5}{c}{\textbf{ESCI-es}} & \multicolumn{5}{c}{\textbf{ESCI-jp}} \\
    \cmidrule(lr){2-6} \cmidrule(lr){7-11} \cmidrule(lr){12-16}
     & R@5 & R@10 & R@100 & N@10 & N@100 & R@5 & R@10 & R@100 & N@10 & N@100 & R@5 & R@10 & R@100 & N@10 & N@100 \\
    \midrule
    \multicolumn{16}{l}{\textit{\textbf{Sparse Retrieval}}} \\
    \quad BM25~(\citeyear{bm25}) & 4.02 & 5.84 & 14.08 & 5.64 & 8.11 & 4.13 & 6.15 & 16.21 & 7.45 & 10.20 & 4.70 & 7.96 & 16.73 & 8.63 & 11.25 \\
    \midrule
    \multicolumn{16}{l}{\textit{\textbf{Dense Retrieval}}} \\
    \quad DPR~(\citeyear{dpr}) & 5.54 & 8.93 & 29.30 & 8.17 & 15.55 & 5.24 & 7.08 & 25.27 & 8.52 & 14.33 & 4.18 & 7.26 & 23.84 & 8.98 & 14.73 \\
    \quad MPNet~(\citeyear{song2020mpnet}) & 2.76 & 4.58 & 15.30 & 4.12 & 9.83 & 2.53 & 4.02 & 13.27 & 5.87 & 8.91 & 1.13 & 1.97 & 5.58 & 1.79 & 2.82 \\
    \quad Sentence-T5~(\citeyear{sentence-t5}) & 4.63 & 6.97 & 24.59 & 6.84 & 11.58 & -- & -- & -- & -- & -- & -- & -- & -- & -- & -- \\
    \quad BGE-m3~(\citeyear{chen2024m3}) & \underline{6.59} & \underline{9.82} & \underline{32.29} & \underline{9.91} & \underline{16.77} & 5.71 & 9.07 & 28.75 & 9.76 & 16.02 & 5.14 & 8.92 & 27.93 & 10.12 & 17.02 \\
    \midrule
    \multicolumn{16}{l}{\textit{\textbf{Generative Retrieval}}} \\
    \quad DSI\textsubscript{naive}~(\citeyear{dsi}) & 0.42 & 1.74 & 2.03 & 0.32 & 0.99 & 0.28 & 0.94 & 1.83 & 0.77 & 1.60 & 0.19 & 0.24 & 1.35 & 0.21 & 1.11 \\
    \quad DSI\textsubscript{semantic}~(\citeyear{dsi}) & 3.74 & 6.02 & 20.69 & 6.24 & 10.60 & 4.27 & 7.13 & 21.63 & 9.92 & 14.23 & 4.08 & 7.40 & 23.90 & 9.15 & 14.58 \\
    \quad TIGER~(\citeyear{rqvae}) & 4.79 & 7.84 & 25.98 & 7.24 & 12.68 & 3.83 & 7.31 & 25.03 & 8.72 & 14.51 & 4.67 & 7.98 & 25.89 & 9.67 & 15.52 \\
    \quad Hi-Gen~(\citeyear{hi-gen}) & 3.13 & 4.97 & 15.91 & 5.25 & 8.55 & 2.97 & 5.65 & 20.23 & 7.06 & 11.85 & 3.36 & 6.40 & 21.66 & 7.73 & 12.86 \\
    \quad LTRGR~(\citeyear{ltrgr}) & 2.88 & 4.71 & 13.70 & 4.48 & 7.82 & 5.13 & 8.23 & 27.92 & 9.26 & 15.42 & 5.24 & 8.37 & 26.79 & 10.01 & 16.93 \\
    \quad RIPOR~(\citeyear{ripor}) & 5.05 & 8.35 & 26.80 & 7.92 & 13.75 & 3.49 & 7.12 & 23.03 & 7.94 & 13.82 & 5.02 & 8.18 & 27.87 & 9.92 & 17.28 \\
    \quad CAT-ID$^2$~(\citeyear{catid}) & 6.11 & 8.89 & 29.03 & 8.96 & 14.53 & 5.70 & \underline{9.34} & \underline{31.44} & \underline{10.68} & \underline{18.01} & \underline{5.36} & 8.87 & \underline{28.82} & 10.34 & 17.14 \\
    \quad MERGE~(\citeyear{merge}) & 5.68 & 9.26 & 29.74 & 9.05 & 15.17 & \underline{5.80} & 9.25 & 30.86 & 10.20 & 17.45 & 5.21 & \underline{8.90} & 27.64 & \underline{10.95} & \underline{17.60} \\
    \midrule
    \midrule
    \quad \method{} & \textbf{7.22}$^*$ & \textbf{11.75}$^*$ & \textbf{36.15}$^*$ & \textbf{10.75}$^*$ & \textbf{18.14}$^*$ & \textbf{6.06}$^*$ & \textbf{10.43}$^*$ & \textbf{34.83}$^*$ & \textbf{12.15}$^*$ & \textbf{20.03}$^*$ & \textbf{5.57}$^*$ & \textbf{9.64}$^*$ & \textbf{31.71}$^*$ & \textbf{11.74}$^*$ & \textbf{18.85}$^*$ \\
    \bottomrule
  \end{tabular}
  }
\end{table}

Table~\ref{tab:performance_comparison} summarizes the experimental results of \method{} across three datasets. \method{} achieves the best performance on all datasets and all evaluation metrics, which shows that the improvement is not limited to a specific language or retrieval depth. Compared with the strongest non-\method{} baseline, \method{} improves R@100 by more than 10\% on both ESCI-us and ESCI-es, and it also keeps clear gains on ESCI-jp. These results suggest that the proposed method is stable under multilingual e-commerce search settings.

The improvement over sparse and dense retrieval baselines indicates that \method{} benefits from modeling more than direct query-item matching. BM25 mainly relies on lexical overlap, while dense retrievers such as DPR, Sentence-T5, MPNet, and BGE-m3 match query and item representations in a shared vector space. These methods can be effective, but short shopping queries, long-tail attributes, and category ambiguity still make the matching problem difficult. \method{} instead learns a category-guided latent intent path before SID generation, which gives the model an intermediate coarse-to-fine understanding of the query. \method{} also outperforms recent generative retrieval baselines, including TIGER, RIPOR, MERGE, and CAT-ID$^2$. Since these methods mainly focus on item SID construction or SID-based retrieval, the gains of \method{} suggest that reducing the semantic gap between natural language queries and abstract SID tokens is also important. The improvements on both recall and NDCG further show that \method{} retrieves more relevant items and ranks them higher, rather than only increasing the size of the retrieved candidate set.

\subsection{Ablation Study (RQ2)}
\begin{table}[htbp]
  \centering
  \setlength\tabcolsep{2.8pt}
  \caption{Ablation study quantifying the contribution of each component.}
  \label{tab:ablation}
  \resizebox{\textwidth}{!}{
    \begin{tabular}{l|c|c|c| ccccc ccccc ccccc}
    \toprule
    \multirow{2}{*}{\textbf{Variant}} & \multicolumn{3}{c}{\textbf{Components}} & \multicolumn{4}{c}{\textbf{ESCI-us}} & \multicolumn{4}{c}{\textbf{ESCI-es}} & \multicolumn{4}{c}{\textbf{ESCI-jp}} \\
    \cmidrule(lr){2-4} \cmidrule(lr){5-8} \cmidrule(lr){9-12} \cmidrule(lr){13-16}
    & HSR & QRE & RCD & R@10 & R@100 & N@10 & N@100 & R@10 & R@100 & N@10 & N@100 & R@10 & R@100 & N@10 & N@100 \\
    \midrule
    TIGER & & & & 7.84 & 25.98 & 7.24 & 12.68 & 7.31 & 25.03 & 8.72 & 14.51 & 7.98 & 25.89 & 9.67 & 15.52 \\
    \midrule
    w/ empty reasoning & & & & 7.81 & 26.38 & 7.30 & 12.84 & 7.25 & 25.17 & 8.80 & 14.44 & 8.05 & 26.01 & 9.84 & 15.65 \\
    w/ HSR & \ding{51} & & & 10.15 & 32.76 & 8.92 & 15.92 & 9.49 & 30.23 & 11.26 & 18.48 & 8.62 & 27.52 & 10.82 & 16.64 \\
    w/ HSR + QRE & \ding{51} & \ding{51} & & 11.62 & 34.10 & 10.65 & 17.57 & 9.60 & 31.44 & 11.48 & 18.55 & 8.72 & 28.37 & 11.25 & 17.32 \\
    \midrule
    \method{} & \ding{51} & \ding{51} & \ding{51} & \textbf{11.75} & \textbf{36.15} & \textbf{10.75} & \textbf{18.14} & \textbf{10.43} & \textbf{34.83} & \textbf{12.15} & \textbf{20.03} & \textbf{9.64} & \textbf{31.71} & \textbf{11.74} & \textbf{18.85} \\
    \bottomrule
    \end{tabular}
  }
\end{table}

Table~\ref{tab:ablation} presents the ablation studies on three datasets to evaluate the contribution of each module in \method{}. The variants include \textit{w/ empty reasoning}, which keeps the reasoning structure but removes supervised signals for latent intent reasoning; \textit{w/ HSR}, which only employs Hierarchical Semantic Reasoning; \textit{w/ HSR + QRE}, which further adds Query-wise Reasoning Enhancement; and \method{}, which builds upon the former by adding Reasoning-aware Constrained Decoding. From the results, we have the following observations:
\begin{itemize}
    \item The \textit{w/ empty reasoning} variant performs close to TIGER across the three datasets. This result shows that simply inserting additional latent states is not enough. Without category-level supervision, the latent states do not receive a clear training signal about how to represent coarse-to-fine shopping intent, so the model cannot reliably benefit from the extra reasoning structure.

    \item Adding HSR brings a clear improvement over both TIGER and \textit{w/ empty reasoning}. This is consistent with the design of HSR: each latent step is directly aligned with a category level, so the decoder receives intermediate intent information before SID generation. The improvement on both recall and NDCG indicates that this category-level supervision helps not only retrieve more relevant items but also rank them higher.

    \item QRE further improves the HSR-only variant. Since one shopping query may correspond to multiple relevant products and categories, a single category label may provide incomplete supervision. QRE uses multi-positive category prototypes under the same query, which gives the latent states a more flexible supervision signal. The gain from \textit{w/ HSR} to \textit{w/ HSR + QRE} supports that modeling multi-positive query intent is useful in e-commerce retrieval.

    \item RCD gives the final improvement from \textit{w/ HSR + QRE} to \method{}. This result is aligned with the inference design: RCD uses the inferred intent categories to assemble a query-specific dynamic prefix trie from pre-indexed category-level tries and prunes invalid SID prefixes during decoding. The candidate space is reduced from the magnitude of $10^5$ to $10^3 \sim 10^4$, which makes the generation process focus on products that are more consistent with the predicted category intent.
\end{itemize}

\subsection{Hyperparameter Analysis (RQ3)}
\subsubsection{Training and Decoding Hyperparameters.}
\begin{figure}[htbp]
  \centering
  \begin{subfigure}[t]{0.32\linewidth}
    \centering
    \includegraphics[width=\linewidth]{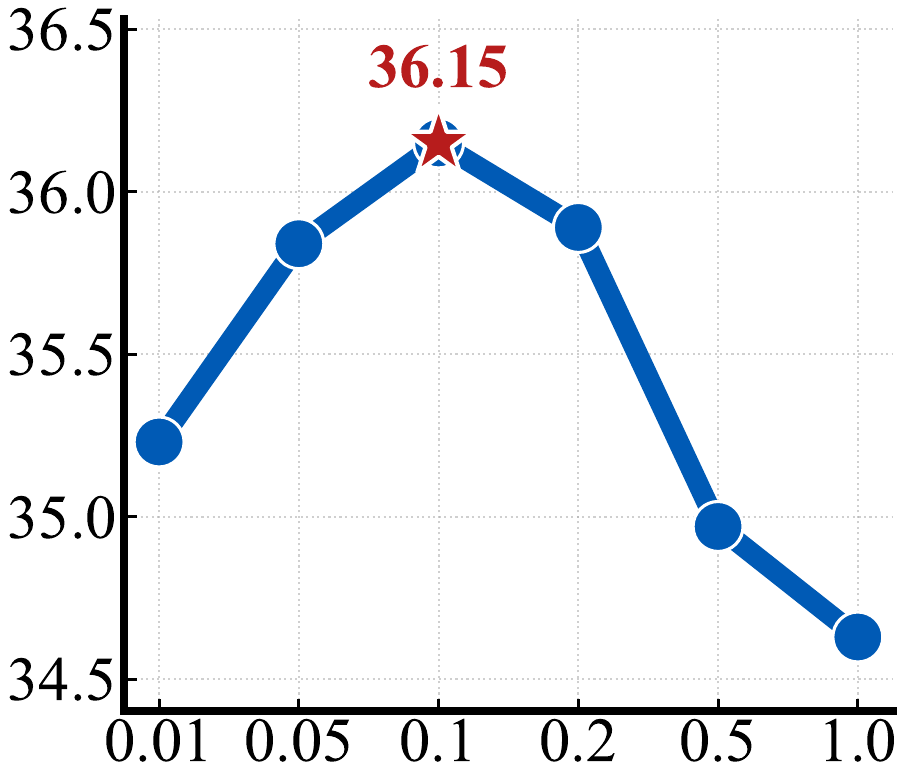}
    \caption{Weight for HSR $\alpha$}
    \label{fig:hparam_sensitivity_us_alpha}
  \end{subfigure}
  \hfill
  \begin{subfigure}[t]{0.32\linewidth}
    \centering
    \includegraphics[width=\linewidth]{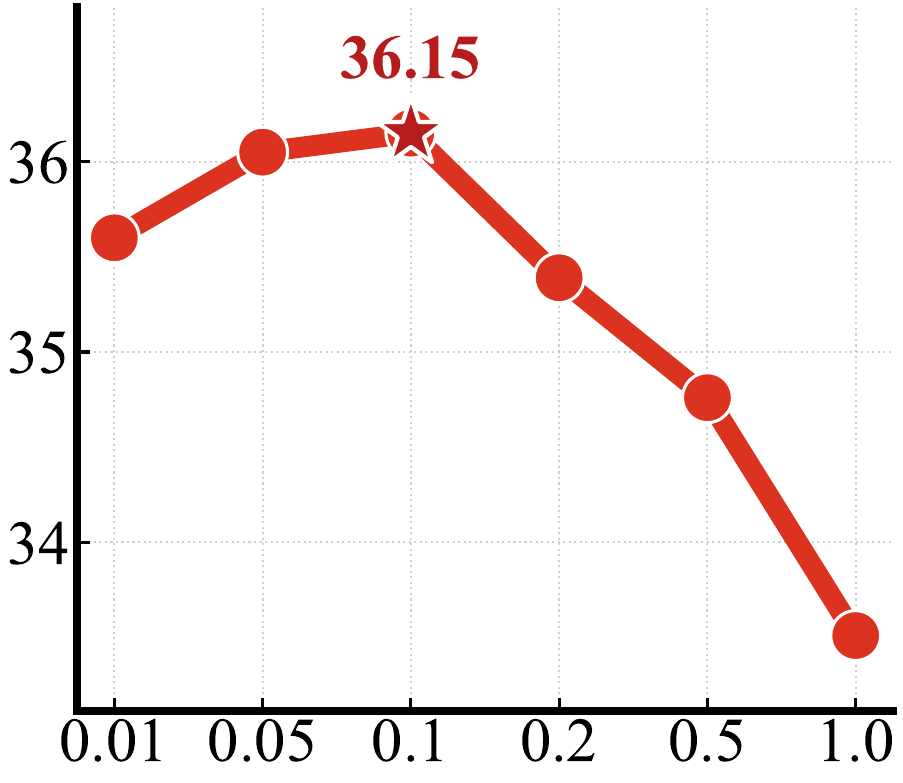}
    \caption{Weight for QRE $\beta$}
    \label{fig:hparam_sensitivity_us_beta}
  \end{subfigure}
  \hfill
  \begin{subfigure}[t]{0.32\linewidth}
    \centering
    \includegraphics[width=\linewidth]{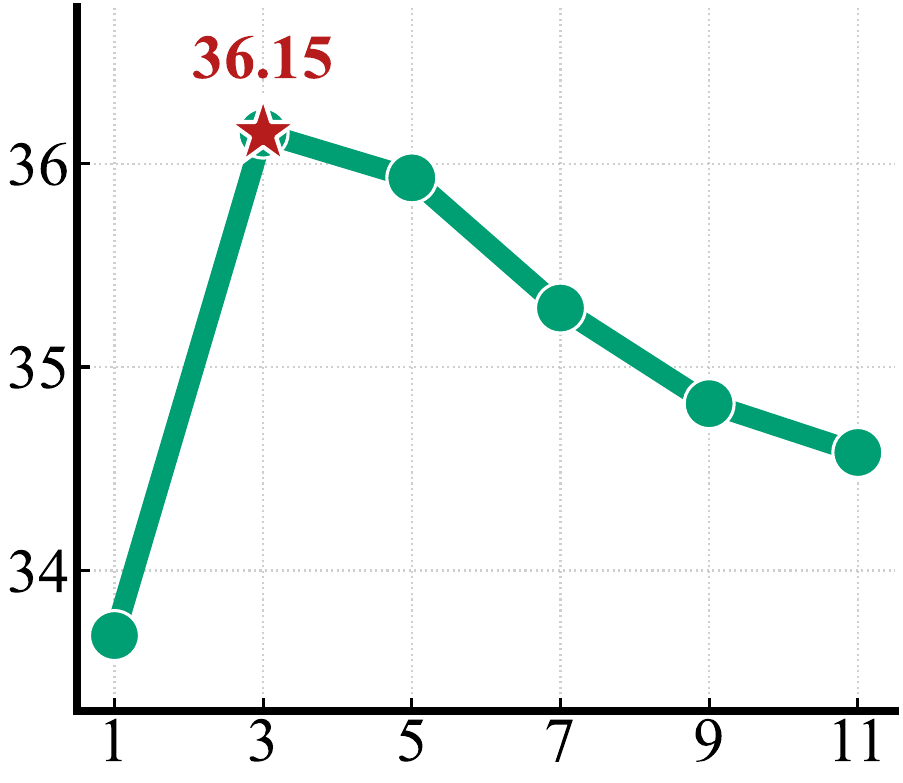}
    \caption{$K$ for Trie}
    \label{fig:hparam_sensitivity_us_K}
  \end{subfigure}
  \caption{Hyperparameter analysis (R@100) on ESCI-us.}
  \label{fig:hparam_sensitivity_us}
\end{figure}

Fig.~\ref{fig:hparam_sensitivity_us} presents the experimental results regarding the three hyperparameters involved in \method{}. Specifically, $\alpha$ controls the learning weight of hierarchical semantic reasoning, $\beta$ balances the importance of query-wise reasoning enhancement, and $K$ denotes the number of categories used for assembling the reasoning-aware dynamic trie. For both $\alpha$ and $\beta$, the performance first increases and then decreases. This trend is consistent with their roles as auxiliary objectives. If the weight is too small, the latent states receive limited category-level guidance; if the weight is too large, the auxiliary objective can compete with the main Seq2Seq generation loss and make the model less focused on final item retrieval. Therefore, HSR and QRE are useful, but they need to be balanced with the SID generation objective.

For the decoding hyperparameter $K$, the best result appears at $K=3$ in the tested setting. A small $K$ may exclude relevant categories from the dynamic trie and thus hurt recall. A larger $K$ includes more categories and more candidate SIDs, but it also introduces more unrelated candidates. This explains why increasing $K$ does not always improve retrieval quality. The result supports the use of a compact but not overly narrow category set during reasoning-aware constrained decoding.

\subsubsection{Indexing Hyperparameters.}
\begin{figure}[htbp]
  \centering
  \begin{subfigure}[t]{0.43\linewidth}
    \centering
    \includegraphics[width=\linewidth]{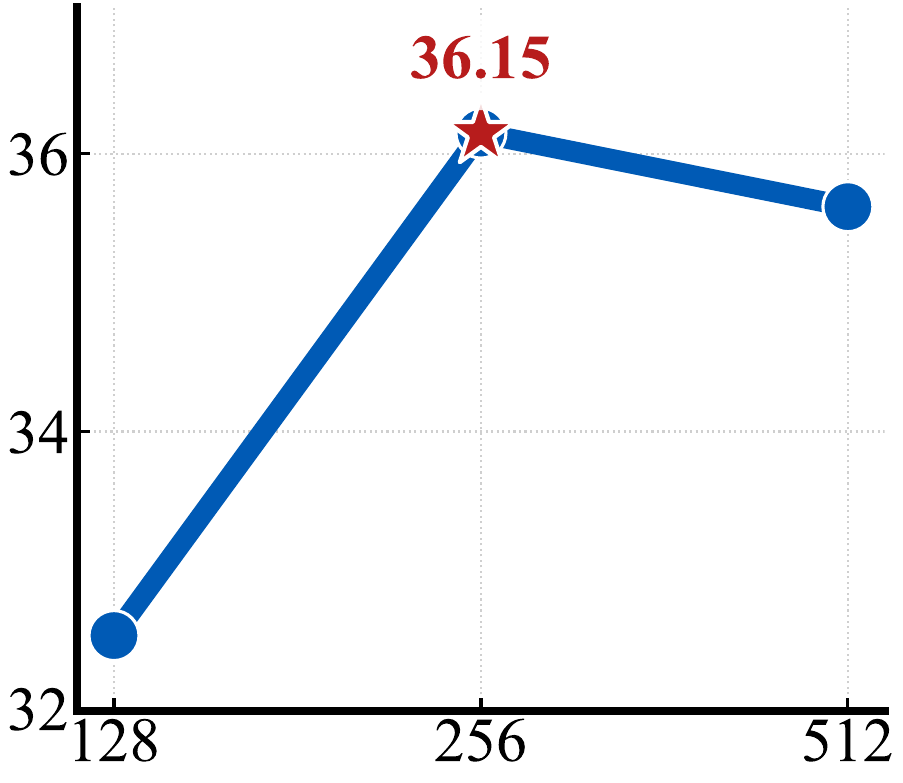}
    \caption{Impact of Codebook Size}
    \label{fig:ablation_codebook}
  \end{subfigure}
  \hfill
  \begin{subfigure}[t]{0.43\linewidth}
    \centering
    \includegraphics[width=\linewidth]{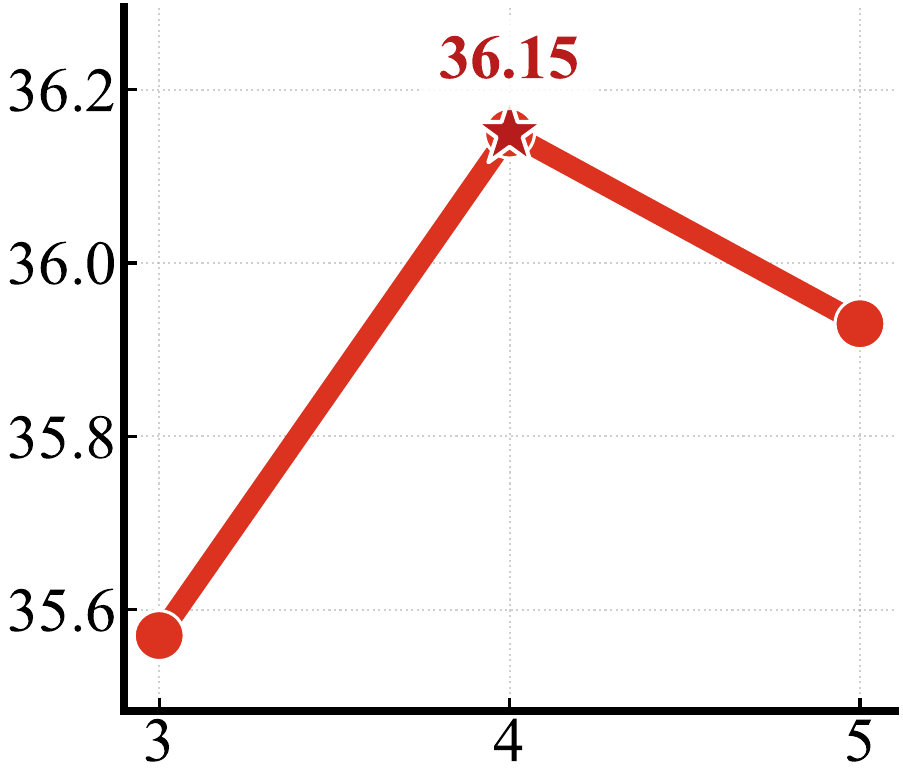}
    \caption{Impact of SID Length}
    \label{fig:ablation_sid}
  \end{subfigure}
  \caption{Parameter analysis on codebook size and SID length over ESCI-us.}
  \label{fig:ablation_experiments}
\end{figure}

We further conduct hyperparameter analysis regarding the Semantic ID (SID) length and codebook size. Fig.~\ref{fig:ablation_experiments} reports the results on ESCI-us. Among the tested settings, a codebook size of 256 gives the best result. A smaller codebook has fewer discrete codes, so different products are more likely to share coarse or less discriminative identifiers. A larger codebook provides more codes, but it may also make the learned SID space less compact. For SID length, the best result appears when the length is 4. A shorter SID may not provide enough levels to distinguish products in a large catalog, while a longer SID adds decoding steps whose later tokens may contribute limited additional discrimination. These results show that \method{} still depends on a reasonably structured item index: latent intent reasoning can guide decoding, but the target SID space itself also needs to preserve enough product-level distinction.

\subsection{Comparison with Explicit CoT/SIDs (RQ4)}
\begin{table}[htbp]
  \centering
  \caption{Comparison with explicit chain-of-thought (CoT) on ESCI-us.}
  \label{tab:cot_comparison_us}
  \begin{tabular}{lccccc}
    \toprule
    \textbf{Model} & R@5 & R@10 & R@100 & N@10 & N@100 \\
    \midrule
    \method{} & \textbf{7.22} & \textbf{11.75} & \textbf{36.15} & \textbf{10.75} & \textbf{18.14} \\
    \midrule
    Explicit CoT & 4.46 & 7.53 & 24.05 & 7.17 & 12.65 \\
    Explicit SIDs & 5.86 & 8.29 & 28.48 & 8.41 & 14.87 \\
    \bottomrule
  \end{tabular}
\end{table}

To further explore the effectiveness of category-guided latent intent reasoning, we conduct a comparative analysis by transforming the latent supervision signals into explicit category text and item descriptions for CoT training. We also replace the explicit reasoning paths with individual category SIDs to investigate the setting where the model directly predicts category SIDs before item SIDs. Table~\ref{tab:cot_comparison_us} reports the results.

\method{} performs better than both explicit variants on all reported metrics, which indicates that making the intermediate reasoning process explicit does not automatically improve e-commerce GR. In this task, the model needs to map short and noisy shopping queries to abstract item SIDs, and extra explicit tokens add more autoregressive prediction steps before final retrieval. Explicit CoT gives the weakest performance among the three methods. One possible reason is that the generated reasoning chain becomes part of the decoding condition. If this chain is incomplete or overly narrow, the following beam search may miss other valid product intents. Explicit SIDs perform better than explicit CoT, suggesting that category-level intermediate signals are useful, but they still lag behind \method{}. This result supports our design choice: \method{} learns the reasoning process before SID generation, but keeps it in continuous latent states rather than verbalizing it as extra output tokens. In this way, the model can use category-aware guidance while avoiding the rigidity and additional decoding cost of explicit reasoning.

\subsection{Compatibility across E-commerce Semantic IDs (RQ5)}
\begin{table}[htbp]
  \centering
  \caption{Compatibility analysis with advanced SID construction methods.}
  \label{tab:scalability_docid}
  \begin{tabular}{lccccc}
    \toprule
    \textbf{Model} & R@5 & R@10 & R@100 & N@10 & N@100 \\
    \midrule
    \method{} & 7.22 & 11.75 & 36.15 & 10.75 & 18.14 \\
    \midrule
    \method{}(MERGE) & \textbf{7.28} & \textbf{12.11} & \textbf{37.20} & \textbf{10.79} & \textbf{18.45} \\
    \method{}(CAT-ID$^2$) & 7.24 & 11.90 & 36.77 & 10.76 & 18.29 \\
    \bottomrule
  \end{tabular}
\end{table}

To verify the compatibility of our proposed framework within e-commerce search, we integrate alternative advanced SID construction methods, such as \text{MERGE}~\cite{merge} and \text{CAT-ID$^2$}~\cite{catid}, into the \method{} architecture. Table~\ref{tab:scalability_docid} summarizes the performance under these settings. Both \method{}(MERGE) and \method{}(CAT-ID$^2$) outperform the default \method{} setting on R@100 and N@100, showing that \method{} can benefit from stronger SID construction methods in the tested setting. This is consistent with the roles of the two components: MERGE and CAT-ID$^2$ mainly improve the item identifier space, while \method{} improves how the query is transformed into category-guided latent intent before decoding. Since these two parts operate at different stages, a better SID space can still be used by \method{}'s reasoning-aware decoding process. At the same time, the gains are not identical for MERGE and CAT-ID$^2$, so the result should be interpreted as evidence of compatibility with advanced SIDs, rather than proof that \method{} will produce the same amount of improvement for every identifier construction method.

\subsection{Impact of Category-Guided Latent Intent Reasoning Steps (RQ6)}
\begin{figure}[htbp]
  \centering
  \begin{subfigure}[t]{0.48\linewidth}
    \centering
    \includegraphics[width=\linewidth,height=0.81\linewidth]{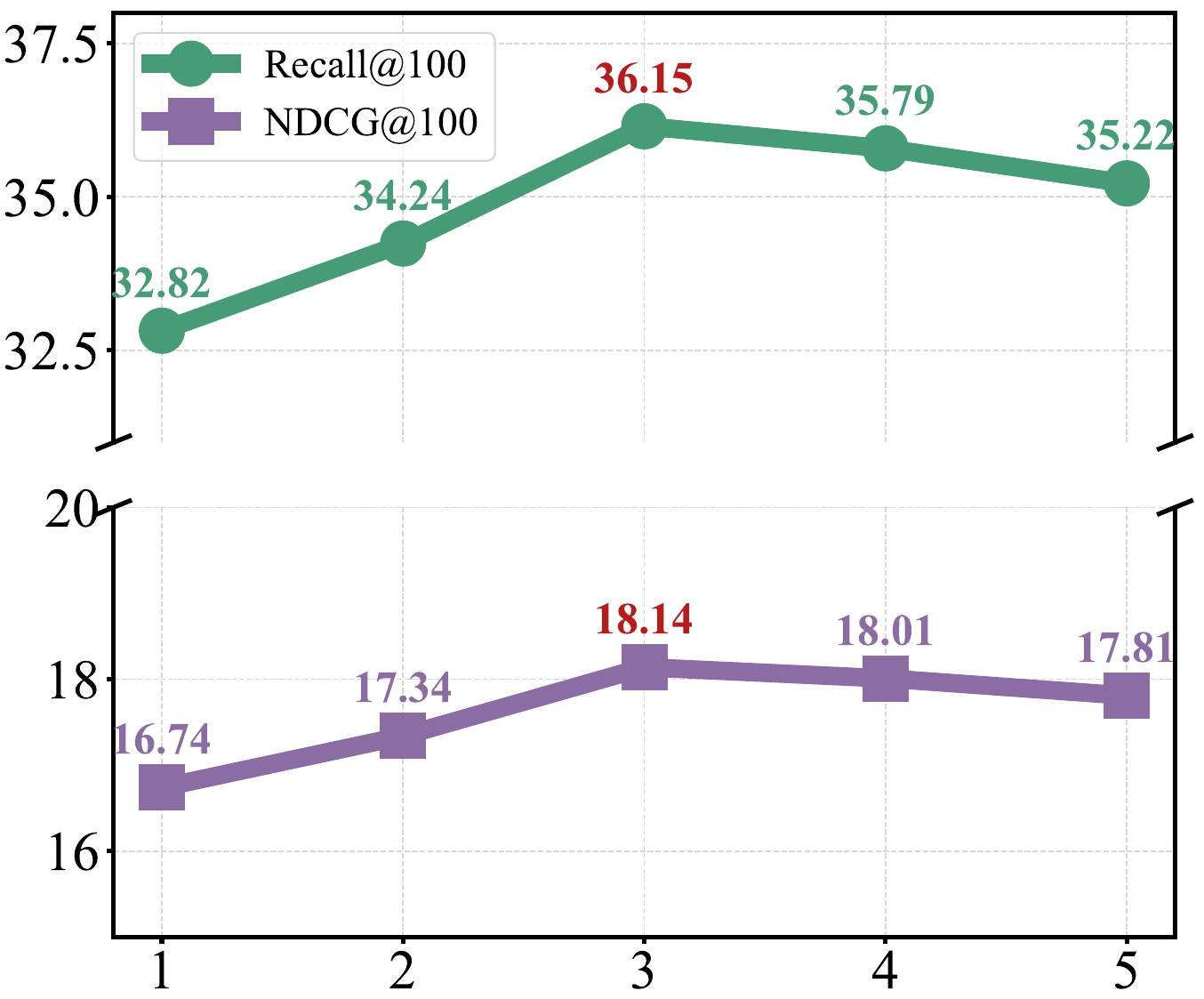}
    \caption{Different Training Steps}
    \label{fig:reasoning_steps_train}
  \end{subfigure}
  \hfill
  \begin{subfigure}[t]{0.48\linewidth}
    \centering
    \includegraphics[width=\linewidth,height=0.81\linewidth]{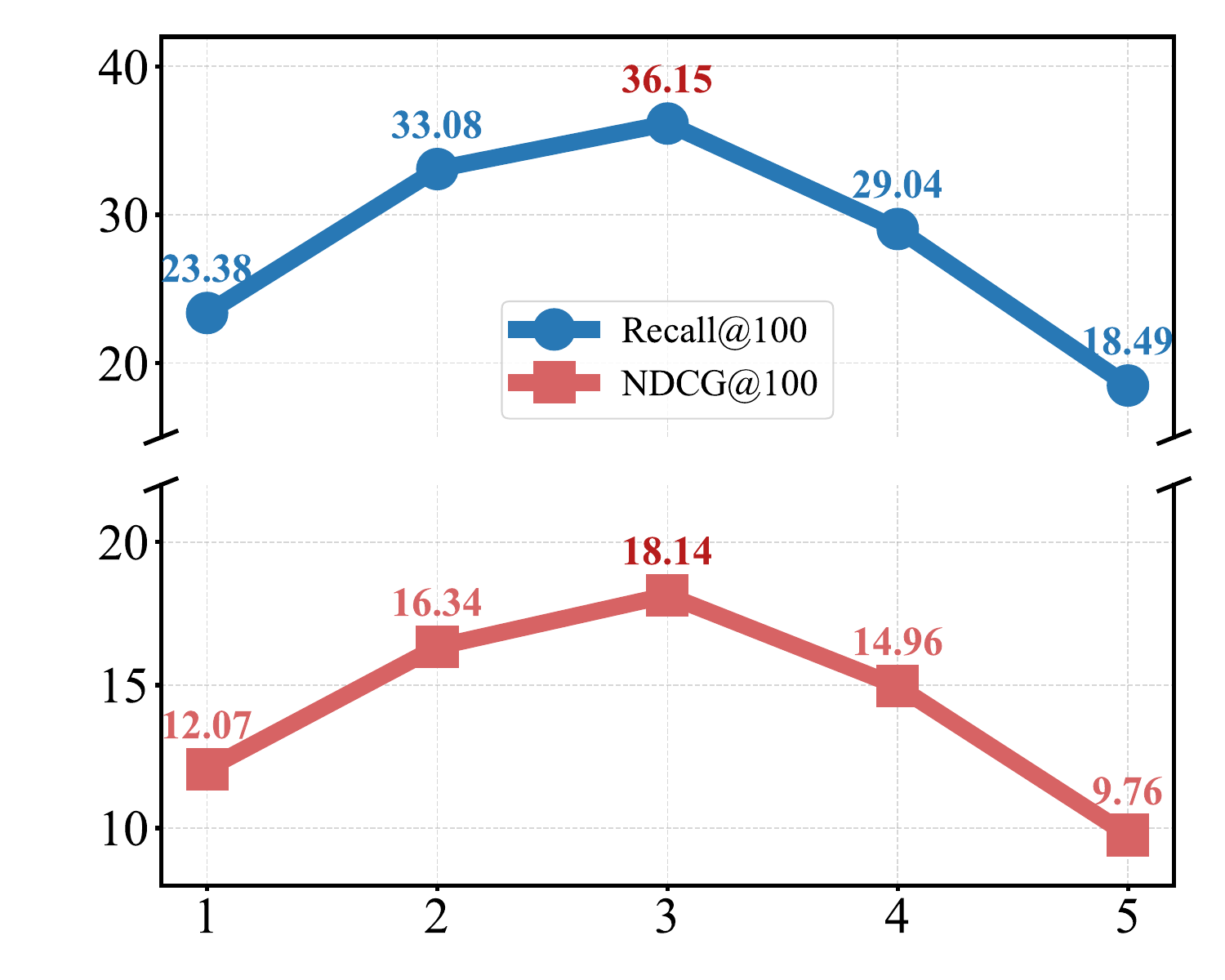}
    \caption{Different Inference Steps}
    \label{fig:reasoning_steps_infer}
  \end{subfigure}
  \caption{Impact of latent reasoning steps on ESCI-us.}
  \label{fig:reasoning_steps_us}
\end{figure}

To further investigate the impact of reasoning steps, which refers to the length of the implicit category-guided intent hierarchy, we conduct two analyses on the ESCI-us dataset. The two subfigures in Fig.~\ref{fig:reasoning_steps_us} have different purposes. Fig.~\ref{fig:reasoning_steps_train} changes the number of reasoning steps during training and is used to choose a fixed reasoning depth offline. In contrast, Fig.~\ref{fig:reasoning_steps_infer} fixes the training step length to 3 and changes the number of reasoning steps during inference, which tests whether the learned reasoning process remains robust when the inference depth is mismatched with the training depth.

During training, the best result appears at 3 reasoning steps. This matches the design of category-guided latent reasoning, where the model learns a coarse-to-fine intent path. With only one or two steps, the latent path has less capacity to represent the full category hierarchy. With more than three steps, the performance slightly decreases, suggesting that simply adding more latent states does not necessarily provide more useful intent information. It is also worth noting that even with a single reasoning step, \method{} still achieves higher R@100 than TIGER in Table~\ref{tab:performance_comparison}. This does not mean one step is optimal, but it shows that a shallow category-guided latent signal can already provide useful guidance beyond direct SID generation.

During inference, using 3 steps gives the best R@100 and NDCG@100 among the tested settings. Reducing the inference steps weakens the latent intent path, while increasing the steps beyond the training setting leads to clear degradation. This setting should not be interpreted as an online search over step numbers. In deployment, \method{} uses the same fixed 3-step configuration selected offline, so it does not introduce per-query step search or additional online tuning cost. The training and inference curves together show that \method{} benefits from a moderate reasoning depth: the model needs enough steps to move from broad categories to fine-grained product groups, but unnecessary extra steps may introduce latent states that are not well aligned with the supervised category hierarchy.

\subsection{Attention and Latent Intent Analysis (RQ7)}
\label{cross-attn}
\begin{figure}[htbp]
  \centering
  \begin{subfigure}[t]{0.45\linewidth}
    \centering
    \includegraphics[width=\linewidth]{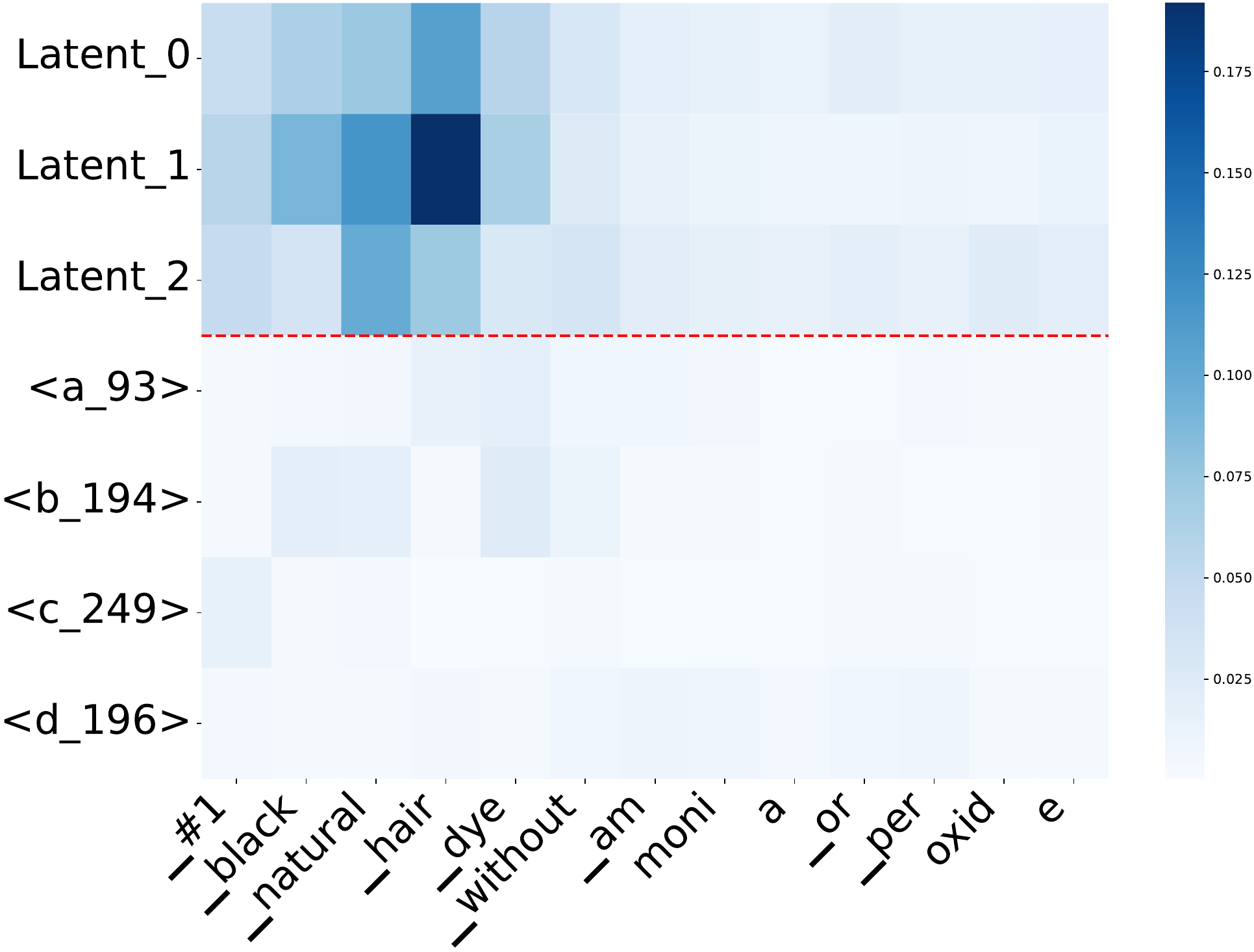}
    \caption{\method{}}
    \label{fig:attention_analysis_calir}
  \end{subfigure}
  \hfill
  \begin{subfigure}[t]{0.48\linewidth}
    \centering
    \includegraphics[width=\linewidth]{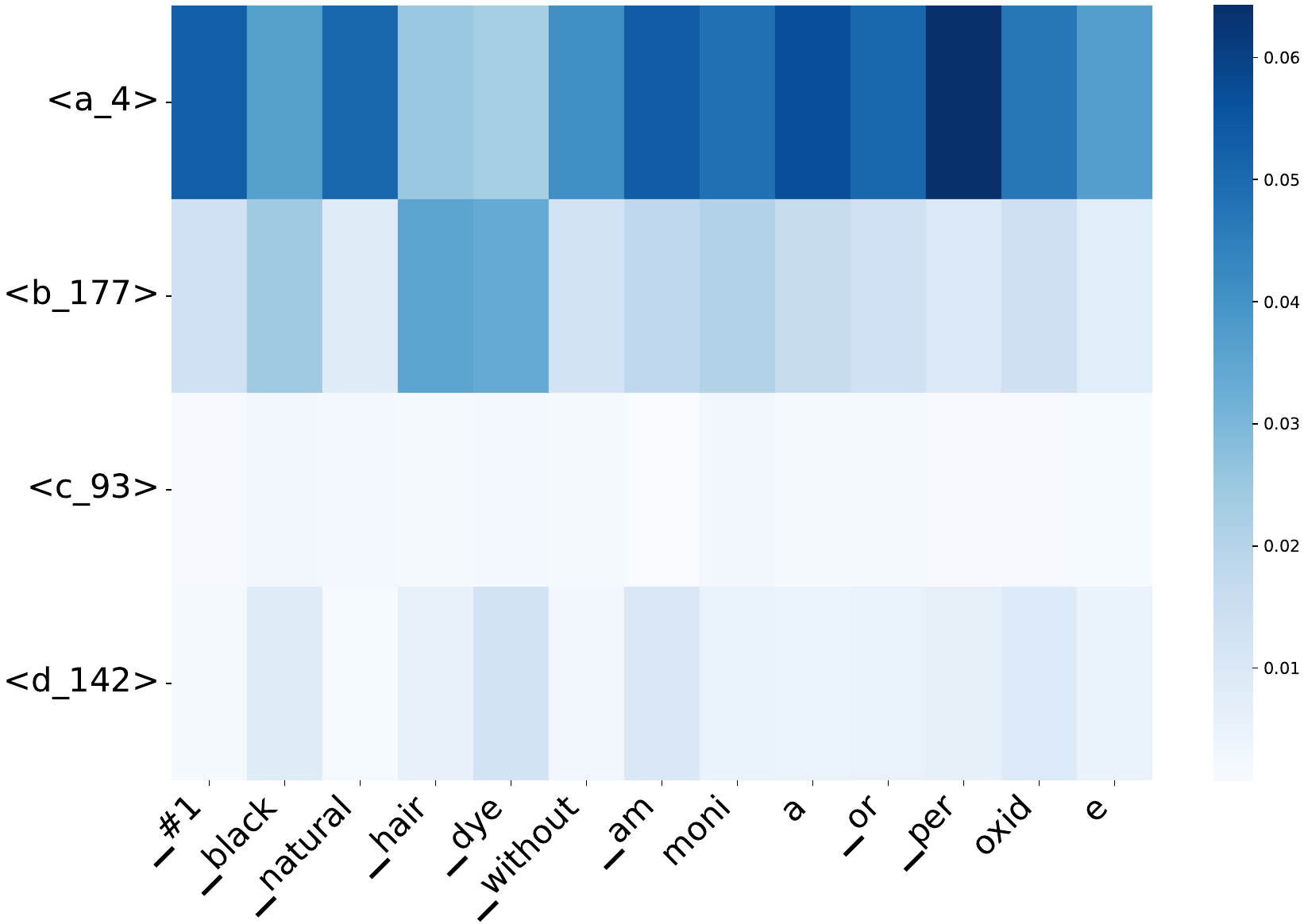}
    \caption{MERGE}
    \label{fig:attention_analysis_merge}
  \end{subfigure}
  \caption{Visualization of cross-attention on ESCI-us.}
  \label{fig:attention_analysis_t5}
\end{figure}

\begin{figure}[htbp]
    \centering
    \includegraphics[width=.6\linewidth]{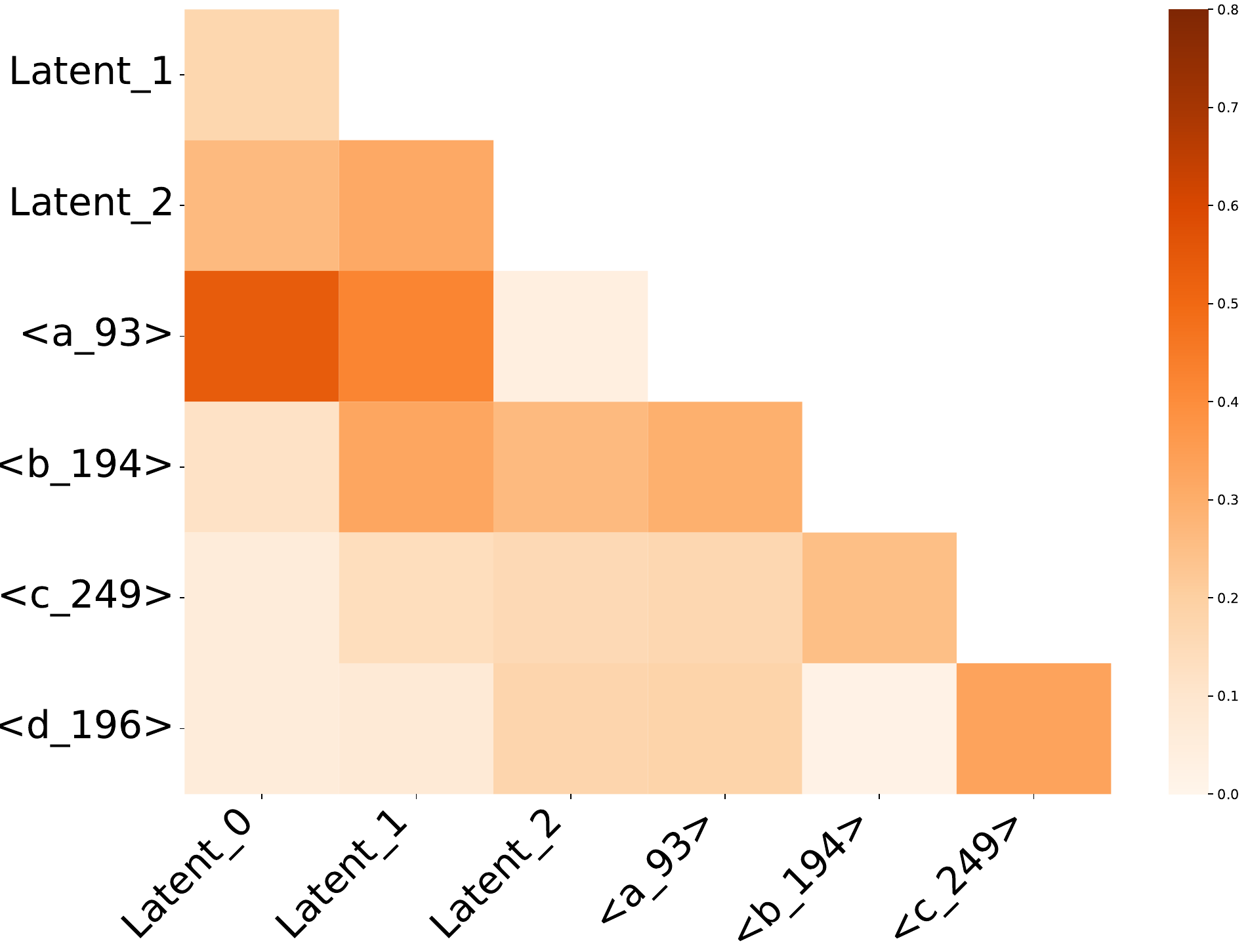}
    \caption{Visualization of self-attention on ESCI-us.}
    \label{fig:self}
\end{figure}

We visualize cross-attention heatmaps for \method{} and MERGE to examine how the decoder uses the query during generation. As shown in Fig.~\ref{fig:attention_analysis_t5}, the latent tokens in \method{} attend to query words such as black, natural, hair, and dye, which are directly related to the product intent in the example query. This pattern suggests that the latent reasoning steps aggregate intent-related evidence from the query before the model starts to generate the final item SID. The generation tokens below the red line show much weaker cross-attention to the original query. This does not mean that the query is ignored. Instead, together with the self-attention visualization in Fig.~\ref{fig:self}, it suggests that the generated SID tokens rely more on the latent states that have already summarized query intent. Compared with \method{}, MERGE shows a more direct attention pattern from generated SID tokens to query words, and the attention scores in this case are less concentrated on a small set of intent words. This provides a concrete example that \method{} separates intent aggregation and SID generation more clearly, while MERGE relies more on direct query-to-SID attention.

Fig.~\ref{fig:self} visualizes the self-attention weights. The generated SID tokens allocate visible attention to the latent hidden states, which provides additional evidence that the latent states are used during SID generation rather than being only auxiliary training signals. This is consistent with the intended reasoning-then-decoding process: the model first forms continuous category-guided intent states, and then the decoder attends to these states when producing the item identifier.

To further analyze whether \method{}'s implicit reasoning captures category-level shopping intent beyond direct query-item matching, we present representative examples in Table~\ref{tab:final_case}. The case studies show the following patterns:
\begin{itemize}
    \item For the long-tail query ``1 1/4'' sink drain without overflow'', \method{} assigns high probability to the correct coarse category path and then separates fine-grained candidates such as ``Bathroom Fixtures'' and ``Kitchen Fixtures''. Since both leaf categories are marked as relevant in the ground truth, this case shows that the model can preserve more than one plausible intent branch instead of forcing an early single-category decision.

    \item For the ambiguous query ``'forget me not i' 2 piece framed print'', \method{} assigns probability mass to both ``Home \& Kitchen'' and ``Handmade Products'' at the root level, and later covers multiple valid leaf categories including ``Posters \& Prints'', ``Artwork'', and ``Paintings''. This behavior is consistent with the multi-positive setting in e-commerce retrieval, where a query can match several category-consistent products.

    \item For the noisy query ``0sxk ofrece fidget toys not expensive'', \method{} still assigns high probability to relevant branches such as ``Toys \& Games'' and finally ``Fidget Toys''. The example suggests that the model can focus on useful shopping-intent words even when the query contains noisy or mixed-language tokens.

    \item These cases are not intended to prove general robustness by themselves, but they provide concrete examples of how the learned latent reasoning path moves from broad product domains to fine-grained product categories. This supports the quantitative results by showing what kind of intermediate intent signal \method{} uses before SID decoding.
\end{itemize}

\begin{table}[htbp]
\centering
\fontsize{6.4pt}{7.2pt}\selectfont
\caption{Case Studies.}
\label{tab:final_case}
\renewcommand{\arraystretch}{1.15}
\setlength{\tabcolsep}{2pt}
\begin{tabularx}{\textwidth}{p{4.35cm} p{2.45cm} X X X}
\toprule
\textbf{Query} & \textbf{Ground Truth} & \textbf{Step 0 (Root)} & \textbf{Step 1 (Level 1)} & \textbf{Step 2 (Leaf)} \\
\midrule
\RaggedRight 1 1/4'' sink drain without overflow &
\RaggedRight
\textbf{L0}: Tools \& Home... \newline
\textbf{L1}: Kitchen \& Bath... \newline
\textbf{L2}: Bathroom Fixt. || Kitchen Fixt. &
\preditem{0.92}{Tools \& Home...}{1} \newline
\preditem{0.04}{Bricolaje y herr...}{0} \newline
\preditem{0.02}{Industrial \& Sci...}{0} \newline
\preditem{0.01}{Power \& Hand...}{0} \newline
\preditem{0.01}{Home \& Kitchen}{0} &
\preditem{0.99}{Kitchen \& Bath Fixt.}{1} \newline
\preditem{0.00}{Accesorios y herr...}{0} \newline
\preditem{0.00}{Rough Plumbing}{0} \newline
\preditem{0.00}{Filtration}{0} \newline
\preditem{0.00}{Kitchen \& Dining}{0} &
\preditem{0.93}{Bathroom Fixtures}{1} \newline
\preditem{0.06}{Kitchen Fixtures}{1} \newline
\preditem{0.00}{Piezas de grifos}{0} \newline
\preditem{0.00}{Faucet Parts}{0} \newline
\preditem{0.00}{Laundry Fixtures}{0}
\\
\midrule
\RaggedRight 'forget me not i' 2 piece framed print &
\RaggedRight
\textbf{L0}: Home \& Kitchen || Handmade Prod. \newline
\textbf{L1}: Wall Art || Home \& Kitchen \newline
\textbf{L2}: Posters \& Prints || Artwork|| Paintings &
\preditem{0.76}{Home \& Kitchen}{1} \newline
\preditem{0.21}{Handmade Prod.}{1} \newline
\preditem{0.01}{Clothing, Shoes...}{0} \newline
\preditem{0.01}{Office Products}{0} \newline
\preditem{0.00}{Baby Products}{0} &
\preditem{0.81}{Wall Art}{1} \newline
\preditem{0.16}{Home \& Kitchen}{1} \newline
\preditem{0.02}{Home Decor Prod...}{0} \newline
\preditem{0.01}{Paint, Wall Treat...}{0} \newline
\preditem{0.00}{Office \& School...}{0} &
\preditem{0.89}{Posters \& Prints}{1} \newline
\preditem{0.07}{Artwork}{1} \newline
\preditem{0.03}{Paintings}{1} \newline
\preditem{0.01}{Photo Albums...}{0} \newline
\preditem{0.00}{Home Decor Acc...}{0}
\\
\midrule
\RaggedRight 0sxk ofrece fidget toys not expensive &
\RaggedRight
\textbf{L0}: Toys || Office|| Home \newline
\textbf{L1}: Novelty|| Baby|| Puzzles|| Party \newline
\textbf{L2}: Fidget|| Squeeze || Miniatures|| Brain Teasers|| Favors &
\preditem{0.65}{Toys \& Games}{1} \newline
\preditem{0.34}{Office Products}{1} \newline
\preditem{0.00}{Health \& Household}{0} \newline
\preditem{0.00}{Sports \& Outdoors}{0} \newline
\preditem{0.00}{Clothing...}{0} &
\preditem{0.70}{Novelty \& Gag...}{1} \newline
\preditem{0.12}{Baby \& Toddler...}{1} \newline
\preditem{0.10}{Party Supplies}{1} \newline
\preditem{0.07}{Puzzles}{1} \newline
\preditem{0.01}{Sports \& Outdoor...}{0} &
\preditem{0.77}{Fidget Toys}{1} \newline
\preditem{0.12}{Miniatures}{1} \newline
\preditem{0.05}{Squeeze Toys}{1} \newline
\preditem{0.03}{Brain Teasers}{1} \newline
\preditem{0.01}{Wind-Up Toys}{0}
\\
\bottomrule
\end{tabularx}
\end{table}

\subsection{Transferability to General-domain Retrieval (RQ8)}
To further examine whether \method{} can be transferred beyond manually defined e-commerce taxonomies, we conduct an additional experiment on the MS MARCO passage-ranking benchmark~\cite{nguyen2016ms}, a general-domain retrieval benchmark with 8.8M passages. Since this setting does not provide product categories or explicit hierarchical labels, we construct a 3-level pseudo hierarchy with 50, 200, and 2000 clusters through hierarchical $K$-means over passage representations. The resulting cluster paths are used only as a coarse-to-fine scaffold for supervising latent reasoning.

\begin{table}[htbp]
  \centering
  \setlength\tabcolsep{8pt}
  \caption{Transferability analysis on MS MARCO.}
  \label{tab:msmarco_transfer}
  \begin{tabular}{lcc}
    \toprule
    \textbf{Model} & \textbf{MRR@10} & \textbf{Recall@10} \\
    \midrule
    LTRGR & 0.255 & 0.402 \\
    RIPOR & 0.333 & 0.562 \\
    CAT-ID$^2$ & 0.338 & 0.533 \\
    \method{}-random & 0.299 & 0.476 \\
    \method{} & \textbf{0.382} & \textbf{0.623} \\
    \bottomrule
  \end{tabular}
\end{table}

Following common MS MARCO passage-ranking practice, we report MRR@10 and additionally report Recall@10 for retrieval coverage. For MRR@10, a query receives the reciprocal rank of the first relevant passage if it appears within the top 10 results, and receives 0 otherwise. Table~\ref{tab:msmarco_transfer} shows that \method{} outperforms the compared GR baselines on both metrics. In this transfer setting, the same induced hierarchy is used wherever a method requires category-like or hierarchical supervision, so \method{} does not use manual taxonomy information unavailable to the baselines. This result suggests that the proposed latent reasoning design can also work when the coarse-to-fine supervision is automatically induced, rather than manually provided by an e-commerce taxonomy.

\method{} also outperforms \method{}-random, which uses the same hierarchy sizes but assigns random labels. This comparison separates the effect of simply adding extra intermediate labels from the effect of using meaningful coarse-to-fine structure. The improvement indicates that the induced hierarchy provides useful supervision for latent reasoning. We emphasize that this experiment should be interpreted as a transferability analysis, not as a change of the main application scenario of this paper. \method{} is still designed and evaluated primarily for e-commerce retrieval, where product categories naturally provide a coarse-to-fine structure. The MS MARCO result only shows that when such taxonomy is unavailable, an automatically constructed hierarchy can provide a usable scaffold for latent reasoning.

\subsection{Backbone Analysis with Qwen (RQ9)}
We further evaluate whether \method{} remains effective when the generative backbone is changed. Specifically, we replace the T5 backbone with Qwen3-0.6B and compare \method{} with TIGER under the same backbone setting. The results are reported in Table~\ref{tab:qwen_backbone}.

\begin{table}[htbp]
  \centering
  \setlength\tabcolsep{5pt}
  \caption{Backbone analysis with Qwen3-0.6B. Each cell reports R@100 / N@10.}
  \label{tab:qwen_backbone}
  \begin{tabular}{lccc}
    \toprule
    \textbf{Model} & \textbf{ESCI-us} & \textbf{ESCI-es} & \textbf{ESCI-jp} \\
    \midrule
    TIGER & 21.59 / 6.95 & 22.04 / 9.44 & 22.66 / 9.35 \\
    \method{} & \textbf{28.37 / 9.31} & \textbf{27.73 / 10.19} & \textbf{27.30 / 10.49} \\
    \bottomrule
  \end{tabular}
\end{table}

Under Qwen3-0.6B, \method{} consistently outperforms TIGER on all three ESCI datasets. Since both methods use the same backbone in this experiment, the improvement is more directly related to the proposed category-guided latent reasoning and reasoning-aware decoding design. The absolute performance under Qwen3-0.6B is lower than the main T5-based setting in Table~\ref{tab:performance_comparison}. Therefore, this experiment is not intended to argue that Qwen3-0.6B is a stronger backbone for our setting. Instead, it checks whether \method{}'s improvement over TIGER still holds when the backbone is replaced. The result suggests that \method{} is not tightly tied to the T5 backbone, and that adding category-guided latent intent reasoning to the GR training and decoding process can still bring gains under a different generative backbone.

\subsection{Inference Efficiency and Latency (RQ10)}
Finally, we analyze the inference latency and parameter scale of \method{}, as summarized in Table~\ref{tab:latency_us}. Latency is measured with batch inference on a single NVIDIA A100 80GB GPU. During inference, \method{} first performs latent intent reasoning for the batch and then runs batched beam search with a query-specific dynamic trie assembled from pre-indexed category-level SID tries. \method{} has higher latency than TIGER and MERGE, which is expected because it performs additional latent reasoning and applies reasoning-aware constrained decoding before final SID generation. However, the increase is moderate in the reported setting, while the R@100 improvement is substantial. Compared with explicit CoT, \method{} is faster and also achieves better R@100, because it does not need to generate a full explicit reasoning chain token by token. \method{} only slightly increases the parameter scale compared with TIGER and MERGE, indicating that the performance gain mainly comes from the added reasoning objectives and decoding strategy rather than from a much larger backbone model. Overall, the results show a practical trade-off: \method{} introduces additional computation over standard GR baselines, but its latency remains below explicit CoT and its retrieval performance is higher in Table~\ref{tab:latency_us}.

\begin{table}[htbp]
  \centering
  \setlength\tabcolsep{4.5pt}
  \caption{Inference time (seconds per query) and model parameters comparison on ESCI-us.}
  \label{tab:latency_us}
  \begin{tabular}{lcccc}
    \toprule
    \textbf{Model} & Inference Time (s) & Params & R@100 \\
    \midrule
    TIGER & 0.258 & 0.223B (1.00x) & 25.98 \\
    MERGE & 0.258 & 0.223B (1.00x) & 29.74 \\
    CoT & 0.433 & 0.223B (1.00x) & 24.05 \\
    \method{} & 0.296 & 0.235B (1.05x) & \textbf{36.15 ($\uparrow$ 21.5\%)} \\
    \bottomrule
  \end{tabular}
\end{table}

\section{Conclusion}
In this paper, we propose \method{}, a category-guided latent intent reasoning framework for generative retrieval~(GR) in e-commerce. \method{} bridges the semantic gap between shopping queries and discrete item SIDs by encoding hierarchical product-category signals into continuous latent intent states.
These states capture salient intent and attribute evidence within the query and serve as a bridge to facilitate the generation of catalog item SIDs. Extensive experiments on multilingual e-commerce search datasets demonstrate that our approach significantly outperforms state-of-the-art baselines, providing a robust and efficient solution for moving beyond direct matching in e-commerce GR. Additional MS MARCO and Qwen3-0.6B analyses further show that the same design remains useful when the hierarchy is automatically induced and when the generative backbone is replaced, while the main focus of this work remains e-commerce retrieval.

\section{Acknowledgments}
This work was supported by the National Key Research and Development Program of China under Grant No. 2024YFF0729003, the National Natural Science Foundation of China under Grant Nos. 62176014, 62206266, and the Fundamental Research Funds for the Central Universities.


\begin{thebibliography}{59}


\ifx \showCODEN    \undefined \def \showCODEN     #1{\unskip}     \fi
\ifx \showDOI      \undefined \def \showDOI       #1{#1}\fi
\ifx \showISBNx    \undefined \def \showISBNx     #1{\unskip}     \fi
\ifx \showISBNxiii \undefined \def \showISBNxiii  #1{\unskip}     \fi
\ifx \showISSN     \undefined \def \showISSN      #1{\unskip}     \fi
\ifx \showLCCN     \undefined \def \showLCCN      #1{\unskip}     \fi
\ifx \shownote     \undefined \def \shownote      #1{#1}          \fi
\ifx \showarticletitle \undefined \def \showarticletitle #1{#1}   \fi
\ifx \showURL      \undefined \def \showURL       {\relax}        \fi
\providecommand\bibfield[2]{#2}
\providecommand\bibinfo[2]{#2}
\providecommand\natexlab[1]{#1}
\providecommand\showeprint[2][]{arXiv:#2}

\bibitem[Bevilacqua et~al\mbox{.}(2022)]%
        {seal}
\bibfield{author}{\bibinfo{person}{Michele Bevilacqua},
  \bibinfo{person}{Giuseppe Ottaviano}, \bibinfo{person}{Patrick Lewis},
  \bibinfo{person}{Scott Yih}, \bibinfo{person}{Sebastian Riedel}, {and}
  \bibinfo{person}{Fabio Petroni}.} \bibinfo{year}{2022}\natexlab{}.
\newblock \showarticletitle{Autoregressive search engines: Generating
  substrings as document identifiers}.
\newblock \bibinfo{journal}{\emph{Advances in Neural Information Processing
  Systems}}  \bibinfo{volume}{35} (\bibinfo{year}{2022}),
  \bibinfo{pages}{31668--31683}.
\newblock


\bibitem[Cao et~al\mbox{.}(2025)]%
        {cao2025onepiece}
\bibfield{author}{\bibinfo{person}{Jiangxia Cao}, \bibinfo{person}{Shuo Yang},
  \bibinfo{person}{Zijun Wang}, {and} \bibinfo{person}{Qinghai Tan}.}
  \bibinfo{year}{2025}\natexlab{}.
\newblock \showarticletitle{OnePiece: The Great Route to Generative
  Recommendation--A Case Study from Tencent Algorithm Competition}.
\newblock \bibinfo{journal}{\emph{arXiv preprint arXiv:2512.07424}}
  (\bibinfo{year}{2025}).
\newblock


\bibitem[Chen et~al\mbox{.}(2024)]%
        {chen2024m3}
\bibfield{author}{\bibinfo{person}{Jianlyu Chen}, \bibinfo{person}{Shitao
  Xiao}, \bibinfo{person}{Peitian Zhang}, \bibinfo{person}{Kun Luo},
  \bibinfo{person}{Defu Lian}, {and} \bibinfo{person}{Zheng Liu}.}
  \bibinfo{year}{2024}\natexlab{}.
\newblock \showarticletitle{M3-embedding: Multi-linguality,
  multi-functionality, multi-granularity text embeddings through self-knowledge
  distillation}. In \bibinfo{booktitle}{\emph{Findings of the Association for
  Computational Linguistics ACL 2024}}. \bibinfo{pages}{2318--2335}.
\newblock


\bibitem[Chen et~al\mbox{.}(2026)]%
        {chen2026lasar}
\bibfield{author}{\bibinfo{person}{Yiwen Chen}, \bibinfo{person}{Fuwei Zhang},
  \bibinfo{person}{Zehao Chen}, \bibinfo{person}{Deqing Wang},
  \bibinfo{person}{Hehan Li}, \bibinfo{person}{Peizhi Xu},
  \bibinfo{person}{Hanmeng Liu}, \bibinfo{person}{Shuanglong Li},
  \bibinfo{person}{Xin Pei}, \bibinfo{person}{Fuzhen Zhuang}, {et~al\mbox{.}}}
  \bibinfo{year}{2026}\natexlab{}.
\newblock \showarticletitle{LASAR: Latent Adaptive Semantic Aligned Reasoning
  for Generative Recommendation}.
\newblock \bibinfo{journal}{\emph{arXiv preprint arXiv:2605.10207}}
  (\bibinfo{year}{2026}).
\newblock


\bibitem[Cheng et~al\mbox{.}(2025)]%
        {cheng2025descriptive}
\bibfield{author}{\bibinfo{person}{Jiehan Cheng}, \bibinfo{person}{Zhicheng
  Dou}, \bibinfo{person}{Yutao Zhu}, {and} \bibinfo{person}{Xiaoxi Li}.}
  \bibinfo{year}{2025}\natexlab{}.
\newblock \showarticletitle{Descriptive and Discriminative Document Identifiers
  for Generative Retrieval}. In \bibinfo{booktitle}{\emph{Proceedings of the
  AAAI Conference on Artificial Intelligence}}, Vol.~\bibinfo{volume}{39}.
  \bibinfo{pages}{11518--11526}.
\newblock


\bibitem[Cuturi(2013)]%
        {cuturi2013sinkhorn}
\bibfield{author}{\bibinfo{person}{Marco Cuturi}.}
  \bibinfo{year}{2013}\natexlab{}.
\newblock \showarticletitle{Sinkhorn distances: Lightspeed computation of
  optimal transport}.
\newblock \bibinfo{journal}{\emph{Advances in neural information processing
  systems}}  \bibinfo{volume}{26} (\bibinfo{year}{2013}).
\newblock


\bibitem[De~Cao et~al\mbox{.}(2021)]%
        {genre}
\bibfield{author}{\bibinfo{person}{Nicola De~Cao}, \bibinfo{person}{Gautier
  Izacard}, \bibinfo{person}{Sebastian Riedel}, {and} \bibinfo{person}{Fabio
  Petroni}.} \bibinfo{year}{2021}\natexlab{}.
\newblock \showarticletitle{Autoregressive Entity Retrieval}. In
  \bibinfo{booktitle}{\emph{International Conference on Learning
  Representations}}. \bibinfo{publisher}{OpenReview.net}.
\newblock


\bibitem[Ferragina and Manzini(2000)]%
        {fmindex}
\bibfield{author}{\bibinfo{person}{Paolo Ferragina} {and}
  \bibinfo{person}{Giovanni Manzini}.} \bibinfo{year}{2000}\natexlab{}.
\newblock \showarticletitle{Opportunistic data structures with applications}.
  In \bibinfo{booktitle}{\emph{Proceedings 41st annual symposium on foundations
  of computer science}}. IEEE, \bibinfo{pages}{390--398}.
\newblock


\bibitem[Hao et~al\mbox{.}(2024)]%
        {coconut}
\bibfield{author}{\bibinfo{person}{Shibo Hao}, \bibinfo{person}{Sainbayar
  Sukhbaatar}, \bibinfo{person}{DiJia Su}, \bibinfo{person}{Xian Li},
  \bibinfo{person}{Zhiting Hu}, \bibinfo{person}{Jason Weston}, {and}
  \bibinfo{person}{Yuandong Tian}.} \bibinfo{year}{2024}\natexlab{}.
\newblock \showarticletitle{Training large language models to reason in a
  continuous latent space}.
\newblock \bibinfo{journal}{\emph{arXiv preprint arXiv:2412.06769}}
  (\bibinfo{year}{2024}).
\newblock


\bibitem[Karpukhin et~al\mbox{.}(2020a)]%
        {dense1}
\bibfield{author}{\bibinfo{person}{Vladimir Karpukhin}, \bibinfo{person}{Barlas
  O{\u{g}}uz}, \bibinfo{person}{Sewon Min}, \bibinfo{person}{Patrick Lewis},
  \bibinfo{person}{Ledell Wu}, \bibinfo{person}{Sergey Edunov},
  \bibinfo{person}{Danqi Chen}, {and} \bibinfo{person}{Wen~Tau Yih}.}
  \bibinfo{year}{2020}\natexlab{a}.
\newblock \showarticletitle{Dense passage retrieval for open-domain question
  answering}. In \bibinfo{booktitle}{\emph{2020 Conference on Empirical Methods
  in Natural Language Processing, EMNLP 2020}}. Association for Computational
  Linguistics (ACL), \bibinfo{pages}{6769--6781}.
\newblock


\bibitem[Karpukhin et~al\mbox{.}(2020b)]%
        {dpr}
\bibfield{author}{\bibinfo{person}{Vladimir Karpukhin}, \bibinfo{person}{Barlas
  Oguz}, \bibinfo{person}{Sewon Min}, \bibinfo{person}{Patrick Lewis},
  \bibinfo{person}{Ledell Wu}, \bibinfo{person}{Sergey Edunov},
  \bibinfo{person}{Danqi Chen}, {and} \bibinfo{person}{Wen-tau Yih}.}
  \bibinfo{year}{2020}\natexlab{b}.
\newblock \showarticletitle{Dense Passage Retrieval for Open-Domain Question
  Answering}. In \bibinfo{booktitle}{\emph{Proceedings of the 2020 Conference
  on Empirical Methods in Natural Language Processing (EMNLP)}}.
  \bibinfo{pages}{6769--6781}.
\newblock


\bibitem[Kuo et~al\mbox{.}(2024)]%
        {kuo2024survey}
\bibfield{author}{\bibinfo{person}{Tzu-Lin Kuo}, \bibinfo{person}{Tzu-Wei
  Chiu}, \bibinfo{person}{Tzung-Sheng Lin}, \bibinfo{person}{Sheng-Yang Wu},
  \bibinfo{person}{Chao-Wei Huang}, {and} \bibinfo{person}{Yun-Nung Chen}.}
  \bibinfo{year}{2024}\natexlab{}.
\newblock \showarticletitle{A survey of generative information retrieval}.
\newblock \bibinfo{journal}{\emph{arXiv preprint arXiv:2406.01197}}
  (\bibinfo{year}{2024}).
\newblock


\bibitem[Li et~al\mbox{.}(2025)]%
        {li2025matching}
\bibfield{author}{\bibinfo{person}{Xiaoxi Li}, \bibinfo{person}{Jiajie Jin},
  \bibinfo{person}{Yujia Zhou}, \bibinfo{person}{Yuyao Zhang},
  \bibinfo{person}{Peitian Zhang}, \bibinfo{person}{Yutao Zhu}, {and}
  \bibinfo{person}{Zhicheng Dou}.} \bibinfo{year}{2025}\natexlab{}.
\newblock \showarticletitle{From matching to generation: A survey on generative
  information retrieval}.
\newblock \bibinfo{journal}{\emph{ACM Trans. Inf. Syst.}} \bibinfo{volume}{43},
  \bibinfo{number}{3} (\bibinfo{year}{2025}), \bibinfo{pages}{1--62}.
\newblock


\bibitem[Li et~al\mbox{.}(2023)]%
        {li-etal-2023-multiview}
\bibfield{author}{\bibinfo{person}{Yongqi Li}, \bibinfo{person}{Nan Yang},
  \bibinfo{person}{Liang Wang}, \bibinfo{person}{Furu Wei}, {and}
  \bibinfo{person}{Wenjie Li}.} \bibinfo{year}{2023}\natexlab{}.
\newblock \showarticletitle{Multiview Identifiers Enhanced Generative
  Retrieval}. In \bibinfo{booktitle}{\emph{Proceedings of the 61st Annual
  Meeting of the Association for Computational Linguistics (Volume 1: Long
  Papers)}}. \bibinfo{publisher}{Association for Computational Linguistics},
  \bibinfo{pages}{6636--6648}.
\newblock


\bibitem[Li et~al\mbox{.}(2024a)]%
        {ltrgr}
\bibfield{author}{\bibinfo{person}{Yongqi Li}, \bibinfo{person}{Nan Yang},
  \bibinfo{person}{Liang Wang}, \bibinfo{person}{Furu Wei}, {and}
  \bibinfo{person}{Wenjie Li}.} \bibinfo{year}{2024}\natexlab{a}.
\newblock \showarticletitle{Learning to rank in generative retrieval}. In
  \bibinfo{booktitle}{\emph{Proceedings of the AAAI Conference on Artificial
  Intelligence}}, Vol.~\bibinfo{volume}{38}. \bibinfo{pages}{8716--8723}.
\newblock


\bibitem[Li et~al\mbox{.}(2024b)]%
        {li2024distillation}
\bibfield{author}{\bibinfo{person}{Yongqi Li}, \bibinfo{person}{Zhen Zhang},
  \bibinfo{person}{Wenjie Wang}, \bibinfo{person}{Liqiang Nie},
  \bibinfo{person}{Wenjie Li}, {and} \bibinfo{person}{Tat-Seng Chua}.}
  \bibinfo{year}{2024}\natexlab{b}.
\newblock \bibinfo{title}{Distillation Enhanced Generative Retrieval}.
\newblock
\newblock
\showeprint[arxiv]{2402.10769}~[cs.CL]


\bibitem[Liu et~al\mbox{.}(2025d)]%
        {liu2025lares}
\bibfield{author}{\bibinfo{person}{Enze Liu}, \bibinfo{person}{Bowen Zheng},
  \bibinfo{person}{Xiaolei Wang}, \bibinfo{person}{Wayne~Xin Zhao},
  \bibinfo{person}{Jinpeng Wang}, \bibinfo{person}{Sheng Chen}, {and}
  \bibinfo{person}{Ji-Rong Wen}.} \bibinfo{year}{2025}\natexlab{d}.
\newblock \showarticletitle{Lares: Latent reasoning for sequential
  recommendation}.
\newblock \bibinfo{journal}{\emph{arXiv preprint arXiv:2505.16865}}
  (\bibinfo{year}{2025}).
\newblock


\bibitem[Liu et~al\mbox{.}(2025a)]%
        {liu2025understanding}
\bibfield{author}{\bibinfo{person}{Jingzhe Liu}, \bibinfo{person}{Liam
  Collins}, \bibinfo{person}{Jiliang Tang}, \bibinfo{person}{Tong Zhao},
  \bibinfo{person}{Neil Shah}, {and} \bibinfo{person}{Clark~Mingxuan Ju}.}
  \bibinfo{year}{2025}\natexlab{a}.
\newblock \showarticletitle{Understanding Generative Recommendation with
  Semantic {IDs} from a Model-Scaling View}.
\newblock \bibinfo{journal}{\emph{arXiv preprint arXiv:2509.25522}}
  (\bibinfo{year}{2025}).
\newblock


\bibitem[Liu et~al\mbox{.}(2025c)]%
        {catid}
\bibfield{author}{\bibinfo{person}{Xiaoyu Liu}, \bibinfo{person}{Fuwei Zhang},
  \bibinfo{person}{Yiqing Wu}, \bibinfo{person}{Xinyu Jia},
  \bibinfo{person}{Zenghua Xia}, \bibinfo{person}{Fuzhen Zhuang},
  \bibinfo{person}{Zhao Zhang}, \bibinfo{person}{Fei Jiang}, {and}
  \bibinfo{person}{Wei Lin}.} \bibinfo{year}{2025}\natexlab{c}.
\newblock \showarticletitle{CAT-ID$^2$: Category-Tree Integrated Document
  Identifier Learning for Generative Retrieval In E-commerce}.
\newblock \bibinfo{journal}{\emph{arXiv preprint arXiv:2511.01461}}
  (\bibinfo{year}{2025}).
\newblock


\bibitem[Liu et~al\mbox{.}(2025b)]%
        {liu2025onerec}
\bibfield{author}{\bibinfo{person}{Zhanyu Liu}, \bibinfo{person}{Shiyao Wang},
  \bibinfo{person}{Xingmei Wang}, \bibinfo{person}{Rongzhou Zhang},
  \bibinfo{person}{Jiaxin Deng}, \bibinfo{person}{Honghui Bao},
  \bibinfo{person}{Jinghao Zhang}, \bibinfo{person}{Wuchao Li},
  \bibinfo{person}{Pengfei Zheng}, \bibinfo{person}{Xiangyu Wu},
  {et~al\mbox{.}}} \bibinfo{year}{2025}\natexlab{b}.
\newblock \showarticletitle{Onerec-think: In-text reasoning for generative
  recommendation}.
\newblock \bibinfo{journal}{\emph{arXiv preprint arXiv:2510.11639}}
  (\bibinfo{year}{2025}).
\newblock


\bibitem[Ma et~al\mbox{.}(2021)]%
        {coil}
\bibfield{author}{\bibinfo{person}{Xinyu Ma}, \bibinfo{person}{Jiafeng Guo},
  \bibinfo{person}{Ruqing Zhang}, \bibinfo{person}{Yixing Fan}, {and}
  \bibinfo{person}{Xueqi Cheng}.} \bibinfo{year}{2021}\natexlab{}.
\newblock \showarticletitle{Contextualized late interaction over dense and
  sparse representations for information retrieval}. In
  \bibinfo{booktitle}{\emph{Proceedings of the 44th International ACM SIGIR
  Conference on Research and Development in Information Retrieval}}.
  \bibinfo{pages}{2401--2405}.
\newblock


\bibitem[Neague et~al\mbox{.}(2024)]%
        {neague2024dsi}
\bibfield{author}{\bibinfo{person}{Petru Neague}, \bibinfo{person}{Marcel
  Gregoriadis}, {and} \bibinfo{person}{Johan Pouwelse}.}
  \bibinfo{year}{2024}\natexlab{}.
\newblock \showarticletitle{De-dsi: Decentralised differentiable search index}.
  In \bibinfo{booktitle}{\emph{Proceedings of the 4th Workshop on Machine
  Learning and Systems}}. \bibinfo{pages}{134--143}.
\newblock


\bibitem[Nguyen et~al\mbox{.}(2016)]%
        {nguyen2016ms}
\bibfield{author}{\bibinfo{person}{Tri Nguyen}, \bibinfo{person}{Mir
  Rosenberg}, \bibinfo{person}{Xia Song}, \bibinfo{person}{Jianfeng Gao},
  \bibinfo{person}{Saurabh Tiwary}, \bibinfo{person}{Rangan Majumder}, {and}
  \bibinfo{person}{Li Deng}.} \bibinfo{year}{2016}\natexlab{}.
\newblock \showarticletitle{Ms marco: A human-generated machine reading
  comprehension dataset}.
\newblock  (\bibinfo{year}{2016}).
\newblock


\bibitem[Ni et~al\mbox{.}(2022)]%
        {sentence-t5}
\bibfield{author}{\bibinfo{person}{Jianmo Ni},
  \bibinfo{person}{Gustavo~Hernandez Abrego}, \bibinfo{person}{Noah Constant},
  \bibinfo{person}{Ji Ma}, \bibinfo{person}{Keith Hall},
  \bibinfo{person}{Daniel Cer}, {and} \bibinfo{person}{Yinfei Yang}.}
  \bibinfo{year}{2022}\natexlab{}.
\newblock \showarticletitle{Sentence-T5: Scalable Sentence Encoders from
  Pre-trained Text-to-Text Models}. In \bibinfo{booktitle}{\emph{Findings of
  the Association for Computational Linguistics: ACL 2022}}.
  \bibinfo{pages}{1864--1874}.
\newblock


\bibitem[Nie et~al\mbox{.}(2025)]%
        {llada}
\bibfield{author}{\bibinfo{person}{Shen Nie}, \bibinfo{person}{Fengqi Zhu},
  \bibinfo{person}{Zebin You}, \bibinfo{person}{Xiaolu Zhang},
  \bibinfo{person}{Jingyang Ou}, \bibinfo{person}{Jun Hu}, \bibinfo{person}{Jun
  Zhou}, \bibinfo{person}{Yankai Lin}, \bibinfo{person}{Ji-Rong Wen}, {and}
  \bibinfo{person}{Chongxuan Li}.} \bibinfo{year}{2025}\natexlab{}.
\newblock \showarticletitle{Large language diffusion models}.
\newblock \bibinfo{journal}{\emph{arXiv preprint arXiv:2502.09992}}
  (\bibinfo{year}{2025}).
\newblock


\bibitem[Rajput et~al\mbox{.}(2023)]%
        {rqvae}
\bibfield{author}{\bibinfo{person}{Shashank Rajput}, \bibinfo{person}{Nikhil
  Mehta}, \bibinfo{person}{Anima Singh}, \bibinfo{person}{Raghunandan
  Hulikal~Keshavan}, \bibinfo{person}{Trung Vu}, \bibinfo{person}{Lukasz
  Heldt}, \bibinfo{person}{Lichan Hong}, \bibinfo{person}{Yi Tay},
  \bibinfo{person}{Vinh Tran}, \bibinfo{person}{Jonah Samost}, {et~al\mbox{.}}}
  \bibinfo{year}{2023}\natexlab{}.
\newblock \showarticletitle{Recommender systems with generative retrieval}.
\newblock \bibinfo{journal}{\emph{Advances in Neural Information Processing
  Systems}}  \bibinfo{volume}{36} (\bibinfo{year}{2023}),
  \bibinfo{pages}{10299--10315}.
\newblock


\bibitem[Reddy et~al\mbox{.}(2022)]%
        {esci}
\bibfield{author}{\bibinfo{person}{Chandan~K Reddy},
  \bibinfo{person}{Llu{\'\i}s M{\`a}rquez}, \bibinfo{person}{Fran Valero},
  \bibinfo{person}{Nikhil Rao}, \bibinfo{person}{Hugo Zaragoza},
  \bibinfo{person}{Sambaran Bandyopadhyay}, \bibinfo{person}{Arnab Biswas},
  \bibinfo{person}{Anlu Xing}, {and} \bibinfo{person}{Karthik Subbian}.}
  \bibinfo{year}{2022}\natexlab{}.
\newblock \showarticletitle{Shopping queries dataset: A large-scale ESCI
  benchmark for improving product search}.
\newblock \bibinfo{journal}{\emph{arXiv preprint arXiv:2206.06588}}
  (\bibinfo{year}{2022}).
\newblock


\bibitem[Reimers and Gurevych(2019)]%
        {reimers2019sentence}
\bibfield{author}{\bibinfo{person}{Nils Reimers} {and} \bibinfo{person}{Iryna
  Gurevych}.} \bibinfo{year}{2019}\natexlab{}.
\newblock \showarticletitle{Sentence-BERT: Sentence Embeddings using Siamese
  BERT-Networks}. In \bibinfo{booktitle}{\emph{Proceedings of the 2019
  Conference on Empirical Methods in Natural Language Processing and the 9th
  International Joint Conference on Natural Language Processing
  (EMNLP-IJCNLP)}}. \bibinfo{publisher}{Association for Computational
  Linguistics}, \bibinfo{pages}{3982--3992}.
\newblock
\urldef\tempurl%
\url{https://doi.org/10.18653/v1/D19-1410}
\showDOI{\tempurl}


\bibitem[Robertson et~al\mbox{.}(2009)]%
        {bm25}
\bibfield{author}{\bibinfo{person}{Stephen Robertson}, \bibinfo{person}{Hugo
  Zaragoza}, {et~al\mbox{.}}} \bibinfo{year}{2009}\natexlab{}.
\newblock \showarticletitle{The probabilistic relevance framework: BM25 and
  beyond}.
\newblock \bibinfo{journal}{\emph{Foundations and Trends{\textregistered} in
  Information Retrieval}} \bibinfo{volume}{3}, \bibinfo{number}{4}
  (\bibinfo{year}{2009}), \bibinfo{pages}{333--389}.
\newblock


\bibitem[Si et~al\mbox{.}(2023)]%
        {seater}
\bibfield{author}{\bibinfo{person}{Zihua Si}, \bibinfo{person}{Zhongxiang Sun},
  \bibinfo{person}{Jiale Chen}, \bibinfo{person}{Guozhang Chen},
  \bibinfo{person}{Xiaoxue Zang}, \bibinfo{person}{Kai Zheng},
  \bibinfo{person}{Yang Song}, \bibinfo{person}{Xiao Zhang},
  \bibinfo{person}{Jun Xu}, {and} \bibinfo{person}{Kun Gai}.}
  \bibinfo{year}{2023}\natexlab{}.
\newblock \showarticletitle{Generative retrieval with semantic tree-structured
  item identifiers via contrastive learning}.
\newblock \bibinfo{journal}{\emph{arXiv preprint arXiv:2309.13375}}
  (\bibinfo{year}{2023}).
\newblock


\bibitem[Song et~al\mbox{.}(2020)]%
        {song2020mpnet}
\bibfield{author}{\bibinfo{person}{Kaitao Song}, \bibinfo{person}{Xu Tan},
  \bibinfo{person}{Tao Qin}, \bibinfo{person}{Jianfeng Lu}, {and}
  \bibinfo{person}{Tie-Yan Liu}.} \bibinfo{year}{2020}\natexlab{}.
\newblock \showarticletitle{Mpnet: Masked and permuted pre-training for
  language understanding}.
\newblock \bibinfo{journal}{\emph{Advances in neural information processing
  systems}}  \bibinfo{volume}{33} (\bibinfo{year}{2020}),
  \bibinfo{pages}{16857--16867}.
\newblock


\bibitem[Sun et~al\mbox{.}(2025)]%
        {sun2025zerogr}
\bibfield{author}{\bibinfo{person}{Weiwei Sun}, \bibinfo{person}{Keyi Kong},
  \bibinfo{person}{Xinyu Ma}, \bibinfo{person}{Shuaiqiang Wang},
  \bibinfo{person}{Dawei Yin}, \bibinfo{person}{Maarten de Rijke},
  \bibinfo{person}{Zhaochun Ren}, {and} \bibinfo{person}{Yiming Yang}.}
  \bibinfo{year}{2025}\natexlab{}.
\newblock \showarticletitle{{ZeroGR}: A Generalizable and Scalable Framework
  for Zero-Shot Generative Retrieval}.
\newblock \bibinfo{journal}{\emph{arXiv preprint arXiv:2510.10419}}
  (\bibinfo{year}{2025}).
\newblock


\bibitem[Sun et~al\mbox{.}(2023)]%
        {genret}
\bibfield{author}{\bibinfo{person}{Weiwei Sun}, \bibinfo{person}{Lingyong Yan},
  \bibinfo{person}{Zheng Chen}, \bibinfo{person}{Shuaiqiang Wang},
  \bibinfo{person}{Haichao Zhu}, \bibinfo{person}{Pengjie Ren},
  \bibinfo{person}{Zhumin Chen}, \bibinfo{person}{Dawei Yin},
  \bibinfo{person}{Maarten Rijke}, {and} \bibinfo{person}{Zhaochun Ren}.}
  \bibinfo{year}{2023}\natexlab{}.
\newblock \showarticletitle{Learning to tokenize for generative retrieval}.
\newblock \bibinfo{journal}{\emph{Advances in Neural Information Processing
  Systems}}  \bibinfo{volume}{36} (\bibinfo{year}{2023}),
  \bibinfo{pages}{46345--46361}.
\newblock


\bibitem[Tang et~al\mbox{.}(2025)]%
        {tang2025think}
\bibfield{author}{\bibinfo{person}{Jiakai Tang}, \bibinfo{person}{Sunhao Dai},
  \bibinfo{person}{Teng Shi}, \bibinfo{person}{Jun Xu}, \bibinfo{person}{Xu
  Chen}, \bibinfo{person}{Wen Chen}, \bibinfo{person}{Jian Wu}, {and}
  \bibinfo{person}{Yuning Jiang}.} \bibinfo{year}{2025}\natexlab{}.
\newblock \showarticletitle{Think before recommend: Unleashing the latent
  reasoning power for sequential recommendation}.
\newblock \bibinfo{journal}{\emph{arXiv preprint arXiv:2503.22675}}
  (\bibinfo{year}{2025}).
\newblock


\bibitem[Tang et~al\mbox{.}(2023)]%
        {sedsi}
\bibfield{author}{\bibinfo{person}{Yubao Tang}, \bibinfo{person}{Ruqing Zhang},
  \bibinfo{person}{Jiafeng Guo}, \bibinfo{person}{Jiangui Chen},
  \bibinfo{person}{Zuowei Zhu}, \bibinfo{person}{Shuaiqiang Wang},
  \bibinfo{person}{Dawei Yin}, {and} \bibinfo{person}{Xueqi Cheng}.}
  \bibinfo{year}{2023}\natexlab{}.
\newblock \showarticletitle{Semantic-enhanced differentiable search index
  inspired by learning strategies}. In \bibinfo{booktitle}{\emph{Proceedings of
  the 29th ACM SIGKDD Conference on Knowledge Discovery and Data Mining}}.
  \bibinfo{pages}{4904--4913}.
\newblock


\bibitem[Tang et~al\mbox{.}(2024)]%
        {grgr}
\bibfield{author}{\bibinfo{person}{Yubao Tang}, \bibinfo{person}{Ruqing Zhang},
  \bibinfo{person}{Jiafeng Guo}, \bibinfo{person}{Maarten de Rijke},
  \bibinfo{person}{Wei Chen}, {and} \bibinfo{person}{Xueqi Cheng}.}
  \bibinfo{year}{2024}\natexlab{}.
\newblock \showarticletitle{Generative Retrieval Meets Multi-Graded Relevance}.
  In \bibinfo{booktitle}{\emph{The Thirty-eighth Annual Conference on Neural
  Information Processing Systems}}.
\newblock


\bibitem[Tay et~al\mbox{.}(2022)]%
        {dsi}
\bibfield{author}{\bibinfo{person}{Yi Tay}, \bibinfo{person}{Vinh Tran},
  \bibinfo{person}{Mostafa Dehghani}, \bibinfo{person}{Jianmo Ni},
  \bibinfo{person}{Dara Bahri}, \bibinfo{person}{Harsh Mehta},
  \bibinfo{person}{Zhen Qin}, \bibinfo{person}{Kai Hui}, \bibinfo{person}{Zhe
  Zhao}, \bibinfo{person}{Jai Gupta}, {et~al\mbox{.}}}
  \bibinfo{year}{2022}\natexlab{}.
\newblock \showarticletitle{Transformer memory as a differentiable search
  index}.
\newblock \bibinfo{journal}{\emph{Advances in Neural Information Processing
  Systems}}  \bibinfo{volume}{35} (\bibinfo{year}{2022}),
  \bibinfo{pages}{21831--21843}.
\newblock


\bibitem[Vaswani et~al\mbox{.}(2017)]%
        {vaswani2017attention}
\bibfield{author}{\bibinfo{person}{Ashish Vaswani}, \bibinfo{person}{Noam
  Shazeer}, \bibinfo{person}{Niki Parmar}, \bibinfo{person}{Jakob Uszkoreit},
  \bibinfo{person}{Llion Jones}, \bibinfo{person}{Aidan~N Gomez},
  \bibinfo{person}{{\L}ukasz Kaiser}, {and} \bibinfo{person}{Illia
  Polosukhin}.} \bibinfo{year}{2017}\natexlab{}.
\newblock \showarticletitle{Attention is all you need}.
\newblock \bibinfo{journal}{\emph{Advances in neural information processing
  systems}}  \bibinfo{volume}{30} (\bibinfo{year}{2017}).
\newblock


\bibitem[Wang et~al\mbox{.}(2024)]%
        {letter}
\bibfield{author}{\bibinfo{person}{Wenjie Wang}, \bibinfo{person}{Honghui Bao},
  \bibinfo{person}{Xinyu Lin}, \bibinfo{person}{Jizhi Zhang},
  \bibinfo{person}{Yongqi Li}, \bibinfo{person}{Fuli Feng},
  \bibinfo{person}{See-Kiong Ng}, {and} \bibinfo{person}{Tat-Seng Chua}.}
  \bibinfo{year}{2024}\natexlab{}.
\newblock \showarticletitle{Learnable item tokenization for generative
  recommendation}. In \bibinfo{booktitle}{\emph{Proceedings of the 33rd ACM
  International Conference on Information and Knowledge Management}}.
  \bibinfo{pages}{2400--2409}.
\newblock


\bibitem[Wang et~al\mbox{.}(2022)]%
        {nci}
\bibfield{author}{\bibinfo{person}{Yujing Wang}, \bibinfo{person}{Yingyan Hou},
  \bibinfo{person}{Haonan Wang}, \bibinfo{person}{Ziming Miao},
  \bibinfo{person}{Shibin Wu}, \bibinfo{person}{Qi Chen},
  \bibinfo{person}{Yuqing Xia}, \bibinfo{person}{Chengmin Chi},
  \bibinfo{person}{Guoshuai Zhao}, \bibinfo{person}{Zheng Liu},
  {et~al\mbox{.}}} \bibinfo{year}{2022}\natexlab{}.
\newblock \showarticletitle{A neural corpus indexer for document retrieval}.
\newblock \bibinfo{journal}{\emph{Advances in Neural Information Processing
  Systems}}  \bibinfo{volume}{35} (\bibinfo{year}{2022}),
  \bibinfo{pages}{25600--25614}.
\newblock


\bibitem[Wei et~al\mbox{.}(2022)]%
        {cot}
\bibfield{author}{\bibinfo{person}{Jason Wei}, \bibinfo{person}{Xuezhi Wang},
  \bibinfo{person}{Dale Schuurmans}, \bibinfo{person}{Maarten Bosma},
  \bibinfo{person}{Fei Xia}, \bibinfo{person}{Ed Chi}, \bibinfo{person}{Quoc~V
  Le}, \bibinfo{person}{Denny Zhou}, {et~al\mbox{.}}}
  \bibinfo{year}{2022}\natexlab{}.
\newblock \showarticletitle{Chain-of-thought prompting elicits reasoning in
  large language models}.
\newblock \bibinfo{journal}{\emph{Advances in neural information processing
  systems}}  \bibinfo{volume}{35} (\bibinfo{year}{2022}),
  \bibinfo{pages}{24824--24837}.
\newblock


\bibitem[Wu et~al\mbox{.}(2025)]%
        {wu2025constrained}
\bibfield{author}{\bibinfo{person}{Shiguang Wu}, \bibinfo{person}{Zhaochun
  Ren}, \bibinfo{person}{Xin Xin}, \bibinfo{person}{Jiyuan Yang},
  \bibinfo{person}{Mengqi Zhang}, \bibinfo{person}{Zhumin Chen},
  \bibinfo{person}{Maarten de Rijke}, {and} \bibinfo{person}{Pengjie Ren}.}
  \bibinfo{year}{2025}\natexlab{}.
\newblock \showarticletitle{Constrained Auto-Regressive Decoding Constrains
  Generative Retrieval}. In \bibinfo{booktitle}{\emph{Proceedings of the 48th
  International ACM SIGIR Conference on Research and Development in Information
  Retrieval}}. \bibinfo{pages}{2429--2440}.
\newblock


\bibitem[Wu et~al\mbox{.}(2024b)]%
        {mvdr}
\bibfield{author}{\bibinfo{person}{Shiguang Wu}, \bibinfo{person}{Wenda Wei},
  \bibinfo{person}{Mengqi Zhang}, \bibinfo{person}{Zhumin Chen},
  \bibinfo{person}{Jun Ma}, \bibinfo{person}{Zhaochun Ren},
  \bibinfo{person}{Maarten de Rijke}, {and} \bibinfo{person}{Pengjie Ren}.}
  \bibinfo{year}{2024}\natexlab{b}.
\newblock \showarticletitle{Generative retrieval as multi-vector dense
  retrieval}. In \bibinfo{booktitle}{\emph{Proceedings of the 47th
  International ACM SIGIR Conference on Research and Development in Information
  Retrieval}}. \bibinfo{pages}{1828--1838}.
\newblock


\bibitem[Wu et~al\mbox{.}(2024a)]%
        {hi-gen}
\bibfield{author}{\bibinfo{person}{Yanjing Wu}, \bibinfo{person}{Yinfu Feng},
  \bibinfo{person}{Jian Wang}, \bibinfo{person}{Wenji Zhou},
  \bibinfo{person}{Yunan Ye}, \bibinfo{person}{Rong Xiao}, {and}
  \bibinfo{person}{Jun Xiao}.} \bibinfo{year}{2024}\natexlab{a}.
\newblock \showarticletitle{Hi-gen: Generative retrieval for large-scale
  personalized e-commerce search}.
\newblock \bibinfo{journal}{\emph{arXiv preprint arXiv:2404.15675}}
  (\bibinfo{year}{2024}).
\newblock


\bibitem[Xiong et~al\mbox{.}(2021)]%
        {dense3}
\bibfield{author}{\bibinfo{person}{Lee Xiong}, \bibinfo{person}{Chenyan Xiong},
  \bibinfo{person}{Ye Li}, \bibinfo{person}{Kwok-Fung Tang},
  \bibinfo{person}{Jialin Liu}, \bibinfo{person}{Paul~N Bennett},
  \bibinfo{person}{Junaid Ahmed}, {and} \bibinfo{person}{Arnold Overwijk}.}
  \bibinfo{year}{2021}\natexlab{}.
\newblock \showarticletitle{Approximate Nearest Neighbor Negative Contrastive
  Learning for Dense Text Retrieval}. In
  \bibinfo{booktitle}{\emph{International Conference on Learning
  Representations}}. \bibinfo{publisher}{OpenReview.net}.
\newblock


\bibitem[Yuan et~al\mbox{.}(2024)]%
        {gdr}
\bibfield{author}{\bibinfo{person}{Peiwen Yuan}, \bibinfo{person}{Xinglin
  Wang}, \bibinfo{person}{Shaoxiong Feng}, \bibinfo{person}{Boyuan Pan},
  \bibinfo{person}{Yiwei Li}, \bibinfo{person}{Heda Wang},
  \bibinfo{person}{Xupeng Miao}, {and} \bibinfo{person}{Kan Li}.}
  \bibinfo{year}{2024}\natexlab{}.
\newblock \showarticletitle{Generative Dense Retrieval: Memory Can Be a
  Burden}. In \bibinfo{booktitle}{\emph{Proceedings of the 18th Conference of
  the European Chapter of the Association for Computational Linguistics (Volume
  1: Long Papers)}}. \bibinfo{pages}{2835--2845}.
\newblock


\bibitem[Zelikman et~al\mbox{.}(2024)]%
        {zelikmanquiet}
\bibfield{author}{\bibinfo{person}{Eric Zelikman}, \bibinfo{person}{Georges
  Harik}, \bibinfo{person}{Yijia Shao}, \bibinfo{person}{Varuna Jayasiri},
  \bibinfo{person}{Nick Haber}, {and} \bibinfo{person}{Noah~D. Goodman}.}
  \bibinfo{year}{2024}\natexlab{}.
\newblock \showarticletitle{Quiet-STaR: Language Models Can Teach Themselves to
  Think Before Speaking}.
\newblock \bibinfo{journal}{\emph{arXiv preprint arXiv:2403.09629}}
  (\bibinfo{year}{2024}).
\newblock


\bibitem[Zeng et~al\mbox{.}(2024)]%
        {ripor}
\bibfield{author}{\bibinfo{person}{Hansi Zeng}, \bibinfo{person}{Chen Luo},
  \bibinfo{person}{Bowen Jin}, \bibinfo{person}{Sheikh~Muhammad Sarwar},
  \bibinfo{person}{Tianxin Wei}, {and} \bibinfo{person}{Hamed Zamani}.}
  \bibinfo{year}{2024}\natexlab{}.
\newblock \showarticletitle{Scalable and effective generative information
  retrieval}. In \bibinfo{booktitle}{\emph{Proceedings of the ACM on Web
  Conference 2024}}. \bibinfo{pages}{1441--1452}.
\newblock


\bibitem[Zhang et~al\mbox{.}(2025a)]%
        {zhang2025hiergr}
\bibfield{author}{\bibinfo{person}{Fuwei Zhang}, \bibinfo{person}{Xiaoyu Liu},
  \bibinfo{person}{Xinyu Jia}, \bibinfo{person}{Yingfei Zhang},
  \bibinfo{person}{Zenghua Xia}, \bibinfo{person}{Fei Jiang},
  \bibinfo{person}{Fuzhen Zhuang}, \bibinfo{person}{Wei Lin}, {and}
  \bibinfo{person}{Zhao Zhang}.} \bibinfo{year}{2025}\natexlab{a}.
\newblock \showarticletitle{HierGR: Hierarchical Semantic Representation
  Enhancement for Generative Retrieval in Food Delivery Search}. In
  \bibinfo{booktitle}{\emph{Proceedings of the 63rd Annual Meeting of the
  Association for Computational Linguistics (Volume 6: Industry Track)}}.
  \bibinfo{pages}{444--455}.
\newblock


\bibitem[Zhang et~al\mbox{.}(2025b)]%
        {merge}
\bibfield{author}{\bibinfo{person}{Fuwei Zhang}, \bibinfo{person}{Xiaoyu Liu},
  \bibinfo{person}{Xinyu Jia}, \bibinfo{person}{Yingfei Zhang},
  \bibinfo{person}{Shuai Zhang}, \bibinfo{person}{Xiang Li},
  \bibinfo{person}{Fuzhen Zhuang}, \bibinfo{person}{Wei Lin}, {and}
  \bibinfo{person}{Zhao Zhang}.} \bibinfo{year}{2025}\natexlab{b}.
\newblock \showarticletitle{Multi-level Relevance Document Identifier Learning
  for Generative Retrieval}. In \bibinfo{booktitle}{\emph{Proceedings of the
  63rd Annual Meeting of the Association for Computational Linguistics (Volume
  1: Long Papers)}}. \bibinfo{pages}{10066--10080}.
\newblock


\bibitem[Zhang et~al\mbox{.}(2026a)]%
        {zhang2026multi}
\bibfield{author}{\bibinfo{person}{Fuwei Zhang}, \bibinfo{person}{Xiaoyu Liu},
  \bibinfo{person}{Dongbo Xi}, \bibinfo{person}{Jishen Yin},
  \bibinfo{person}{Huan Chen}, \bibinfo{person}{Peng Yan},
  \bibinfo{person}{Fuzhen Zhuang}, {and} \bibinfo{person}{Zhao Zhang}.}
  \bibinfo{year}{2026}\natexlab{a}.
\newblock \showarticletitle{Multi-aspect cross-modal quantization for
  generative recommendation}. In \bibinfo{booktitle}{\emph{Proceedings of the
  AAAI Conference on Artificial Intelligence}}, Vol.~\bibinfo{volume}{40}.
  \bibinfo{pages}{16271--16279}.
\newblock


\bibitem[Zhang et~al\mbox{.}(2022)]%
        {zhang2022mind}
\bibfield{author}{\bibinfo{person}{Fuwei Zhang}, \bibinfo{person}{Zhao Zhang},
  \bibinfo{person}{Xiang Ao}, \bibinfo{person}{Dehong Gao},
  \bibinfo{person}{Fuzhen Zhuang}, \bibinfo{person}{Yi Wei}, {and}
  \bibinfo{person}{Qing He}.} \bibinfo{year}{2022}\natexlab{}.
\newblock \showarticletitle{Mind the gap: Cross-lingual information retrieval
  with hierarchical knowledge enhancement}. In
  \bibinfo{booktitle}{\emph{Proceedings of the AAAI Conference on Artificial
  Intelligence}}, Vol.~\bibinfo{volume}{36}. \bibinfo{pages}{4345--4353}.
\newblock


\bibitem[Zhang et~al\mbox{.}(2025d)]%
        {zhang2025slow}
\bibfield{author}{\bibinfo{person}{Junjie Zhang}, \bibinfo{person}{Beichen
  Zhang}, \bibinfo{person}{Wenqi Sun}, \bibinfo{person}{Hongyu Lu},
  \bibinfo{person}{Wayne~Xin Zhao}, \bibinfo{person}{Yu Chen}, {and}
  \bibinfo{person}{Ji-Rong Wen}.} \bibinfo{year}{2025}\natexlab{d}.
\newblock \showarticletitle{Slow thinking for sequential recommendation}.
\newblock \bibinfo{journal}{\emph{arXiv preprint arXiv:2504.09627}}
  (\bibinfo{year}{2025}).
\newblock


\bibitem[Zhang et~al\mbox{.}(2024)]%
        {zhang2024ai}
\bibfield{author}{\bibinfo{person}{Yongfeng Zhang}, \bibinfo{person}{Zhiwei
  Liu}, \bibinfo{person}{Qingsong Wen}, \bibinfo{person}{Linsey Pang},
  \bibinfo{person}{Wei Liu}, {and} \bibinfo{person}{Philip~S Yu}.}
  \bibinfo{year}{2024}\natexlab{}.
\newblock \showarticletitle{{AI} Agent for Information Retrieval: Generating
  and Ranking}. In \bibinfo{booktitle}{\emph{Proceedings of the 33rd ACM
  International Conference on Information and Knowledge Management}}.
  \bibinfo{pages}{5605--5607}.
\newblock


\bibitem[Zhang et~al\mbox{.}(2025c)]%
        {zhang2025replication}
\bibfield{author}{\bibinfo{person}{Zhen Zhang}, \bibinfo{person}{Xinyu Ma},
  \bibinfo{person}{Weiwei Sun}, \bibinfo{person}{Pengjie Ren},
  \bibinfo{person}{Zhumin Chen}, \bibinfo{person}{Shuaiqiang Wang},
  \bibinfo{person}{Dawei Yin}, \bibinfo{person}{Maarten de Rijke}, {and}
  \bibinfo{person}{Zhaochun Ren}.} \bibinfo{year}{2025}\natexlab{c}.
\newblock \showarticletitle{Replication and Exploration of Generative Retrieval
  over Dynamic Corpora}. In \bibinfo{booktitle}{\emph{Proceedings of the 48th
  International ACM SIGIR Conference on Research and Development in Information
  Retrieval}}. \bibinfo{pages}{3325--3334}.
\newblock


\bibitem[Zhang et~al\mbox{.}(2026b)]%
        {zhang2026model}
\bibfield{author}{\bibinfo{person}{Zhen Zhang}, \bibinfo{person}{Zihan Wang},
  \bibinfo{person}{Xinyu Ma}, \bibinfo{person}{Shuaiqiang Wang},
  \bibinfo{person}{Dawei Yin}, \bibinfo{person}{Xin Xin},
  \bibinfo{person}{Pengjie Ren}, \bibinfo{person}{Maarten de Rijke}, {and}
  \bibinfo{person}{Zhaochun Ren}.} \bibinfo{year}{2026}\natexlab{b}.
\newblock \showarticletitle{Model Editing for New Document Integration in
  Generative Information Retrieval}. In \bibinfo{booktitle}{\emph{Proceedings
  of the ACM Web Conference 2026}}. \bibinfo{pages}{1993--2003}.
\newblock


\bibitem[Zhou et~al\mbox{.}(2022a)]%
        {learnedsparse}
\bibfield{author}{\bibinfo{person}{Yujia Zhou}, \bibinfo{person}{Zhicheng Dou},
  {and} \bibinfo{person}{Ji-Rong Wen}.} \bibinfo{year}{2022}\natexlab{a}.
\newblock \showarticletitle{Learning sparse representations for end-to-end
  dense retrieval}. In \bibinfo{booktitle}{\emph{Proceedings of the 45th
  International ACM SIGIR Conference on Research and Development in Information
  Retrieval}}. \bibinfo{pages}{2359--2364}.
\newblock


\bibitem[Zhou et~al\mbox{.}(2022b)]%
        {zhou2022ultron}
\bibfield{author}{\bibinfo{person}{Yujia Zhou}, \bibinfo{person}{Jing Yao},
  \bibinfo{person}{Zhicheng Dou}, \bibinfo{person}{Ledell Wu},
  \bibinfo{person}{Peitian Zhang}, {and} \bibinfo{person}{Ji-Rong Wen}.}
  \bibinfo{year}{2022}\natexlab{b}.
\newblock \showarticletitle{Ultron: An ultimate retriever on corpus with a
  model-based indexer}.
\newblock \bibinfo{journal}{\emph{arXiv preprint arXiv:2208.09257}}
  (\bibinfo{year}{2022}).
\newblock


\bibitem[Zhuang et~al\mbox{.}(2022)]%
        {dsiqg}
\bibfield{author}{\bibinfo{person}{Shengyao Zhuang}, \bibinfo{person}{Houxing
  Ren}, \bibinfo{person}{Linjun Shou}, \bibinfo{person}{Jian Pei},
  \bibinfo{person}{Ming Gong}, \bibinfo{person}{Guido Zuccon}, {and}
  \bibinfo{person}{Daxin Jiang}.} \bibinfo{year}{2022}\natexlab{}.
\newblock \showarticletitle{Bridging the Gap Between Indexing and Retrieval for
  Differentiable Search Index with Query Generation}.
\newblock \bibinfo{journal}{\emph{arXiv preprint arXiv:2206.10128}}
  (\bibinfo{year}{2022}).
\newblock


\end{thebibliography}
\end{document}